# Comparisons of Australian Mental Health Distributions

David Gunawan
*University of Wollongong*

William Griffiths*
*University of Melbourne*

Duangkamon Chotikapanich
*Monash University*

10 June, 2021

**Abstract**

Bayesian nonparametric estimates of Australian mental health distributions are obtained to assess how the mental health status of the population has changed over time, and to compare the mental health status of female/male and indigenous/non-indigenous population subgroups. First- and second-order stochastic dominance are used to compare distributions, with results presented in terms of the posterior probability of dominance and the posterior probability of no dominance. Our results suggest mental health has deteriorated in recent years, that males' mental health status is better than that of females, and non-indigenous health status is better than that of the indigenous population.

*Corresponding author
William Griffiths
Department of Economics
University of Melbourne
Vic 3010
Australia
wegrif@unimelb.edu.au

## 1. Introduction

Improving the general level of health and reducing health inequality are major objectives of public policy. To assess whether improvements are being made overtime, and in different subgroups of a population, we need to sample the health status of individuals from the populations of interest, and to use those samples to make inferences about the populations. These inferences could take the form of comparing health status at different points in time, or comparing the health status of different segments of the population. Such comparisons involve several nontrivial steps that include sampling, measuring health status, choosing criteria for making comparisons, and making inferences about those criteria.

We focus on the mental health status of the Australian population in the years 2001, 2006, 2010, 2014 and 2017, and on that for male/female and indigenous/non-indigenous population subgroups in the same years. The prevalence of poor mental health has attracted increasing attention in recent years, particularly in relation to difficulties resulting from COVID-19 lockdowns. Our sample does not cover the post-COVID period, but our results suggest mental health had already been deteriorating prior to that time. Interest has also centred on the status of female mental health relative to that of males, and a major government policy objective has been to narrow the gap between indigenous and non-indigenous health status. We examine evidence on the relative health status of these population subgroups and how this evidence has changed over time. The sample we use is the SF-36 health survey questions administered as part of the Household, Income and Labour Dynamics in Australia (HILDA) survey.[1] Responses to the mental health questions are converted to a continuous score that ranges between 0 and 100, with 100 representing good mental health. Several criteria are used to compare scores. The novelty in our approach is the use of stochastic dominance as one of the criteria, and the use of Bayesian inference to assess dominance and to estimate other criteria. Bayesian nonparametric methods are used to estimate a distribution of scores from each sample. In addition to comparing mean scores using their respective posterior distributions, we find posterior probabilities of dominance for each pairwise comparison of distributions. Both first and second order stochastic dominance are considered.

---

[1] The HILDA project (Watson and Wooden, 2012) was initiated and is funded by the Australian Government Department of Social Services (DSS) and is managed by the Melbourne Institute of Applied Economics and Social Research (Melbourne Institute). The findings and views in this paper, however, are those of the authors and should not be attributed to either DSS or the Melbourne Institute.

Checking to see if one mental health distribution dominates another involves comparing the cumulative distribution functions (cdf's) of the two sets of health scores. For first-order stochastic dominance (FSD), a distribution *A* dominates a distribution *B* (written $A_{FSD}B$) if the cdf for *A* lies below the cdf for *B*; the proportion of population with a mental health score below any value *y* is less in *A* that it is in *B*. The two cdf's do not cross. For second-order stochastic dominance (SSD), *A* dominates *B* (written $A_{SSD}B$) if the area under the cdf between zero and any value *y* is less for *A*, than it is for *B*. It implies the sum of all mental health scores less than any value *y* is less for *A* than it is for *B*.[2] The existence of SSD does not rule out the possibility of cdf's that cross; FSD implies SSD, but the converse is not true. We present evidence on the relative standing of *A* and *B* in terms of three posterior probabilities: the probability *A* dominates *B*, the probability *B* dominates *A*, and the probability that neither distribution is dominant.

Likely to be of particular concern is the prevalence of severe mental illness – those in the left tail of the distribution. To provide evidence on these tails, we borrow concepts from poverty analysis, providing posterior information on the proportion of the population below a threshold and the severity of illness below that threshold. We also examine dominance results for those in the lower tail of the distribution.

In Section 2 we briefly review examples of studies that have converted categorical data from self-reported health surveys into a continuous variable, and then describe the questionnaire used to obtain observations that are converted into the continuous mental health variable used in this study. The observations on this variable are modelled nonparametrically using an infinite mixture of beta distributions. The Bayesian Markov chain Monte Carlo (MCMC) methodology for estimating this model, and the quantities necessary for comparing mental health distributions, is described in a Supplementary Appendix. The criteria used for comparing distributions are described in Sections 3 and 4, with Section 3 being devoted to single characteristics of the distributions, and Section 4 covering dominance criteria which involve comparing whole distributions. The mean is the single characteristic used for comparing complete distributions. For comparing the left tails of the distributions, where mental health is particularly poor, we use the headcount ratio and two values from an “FGT-family” of measures, concepts borrowed from poverty analysis. In Section 4, we define first- and second-order stochastic dominance, explain how MCMC draws are used to estimate probabilities of dominance, and describe how this approach differs from sampling theory tests for dominance. Our results are presented in Section 5. Briefly, we find poor mental health is more prevalent in 2017 than in 2001, more prevalent in females than males, and more prevalent in indigenous than in non-indigenous subgroups of the population. Some concluding remarks are provided in Section 6.

---

[2] More precise definitions of FSD and SSD that include strict and non-strict inequalities are provided in Section 4.

## 2. Specifying the Health Distribution

Constructing a suitable health distribution is not a simple task; various proposals put forward in the literature all have their disadvantages. A commonly used indicator is self-reported health status (SRHS), typically measured on a scale consisting of three or five categories: good, fair, and poor, or excellent, very good, good, fair, and poor. Categorical measures of health such as these create a problem for the measurement of inequalities and deprivations. They prevent the direct use of traditional tools of distribution analysis, such as the Lorenz curve, the Gini coefficient, and the concentration index, which require continuous variables for evaluating inequality (Wagstaff and van Doorslare, 1994). Several transformations and scaling assumptions have been proposed to cardinalize SRHS. For example, it can be transformed into a discrete variable by assigning arbitrary numerical values (1, 2, 3, 4, 5) to the categories while maintaining the ordered nature of the health responses. This approach can be problematic, however, since small variations in the numerical scale may reverse the ordering of a given pair of distributions. Allison and Foster (2004) found that conventional inequality indices (such as variance, the Gini coefficient, and those developed for income distributions) are sensitive to the particular scale employed to convert health responses into discrete numerical values. Wagstaff and van Doorslaer (1994) proposed a method that assumes the underlying categorical empirical distribution of the responses to SRHS is a latent, continuous, but unobservable health variable with a standard lognormal distribution. A similar scaling approach was adopted by Chotikapanich et al. (2003) to measure the extent of income-related health inequality in Australia. Van Doorslaer and Jones (2003) estimate a continuous latent variable for health by specifying ordered probit and interval regression models in which SRHS is the dependent variable and the latent variable is a function of some socioeconomic and demographic characteristic. They use the predicted health level as a measure of individual health after rescaling, and assess the performance of different methods.

We follow the lead of these researchers in the sense that we represent the distribution of mental health as a continuous variable to facilitate further analysis, and to enable comparison of distributions over time, and for different population subgroups. The HILDA survey, from which we extract the responses to the SF-36 health survey questions, is a national representative longitudinal survey which began in Australia in 2001, and is conducted annually. It collects key variables concerning family and household structure, as well as data on education, income, health, life satisfaction and other variables relating to economic and subjective wellbeing. The SF-36 survey (Medical Outcome Trust, Boston, MA) is a multipurpose and short form health survey with 36 questions that provide one of the most widely used generic and continuous measures of health-related quality of life in clinical research and general population health. It has been translated and studied in more than 40 countries (Ware et al. 1993, Ware and Gandek 1998). The developers of the SF-36 claim that scaling assumptions used to transform the ordered categorical responses into a continuous health measure can be interpreted as “quasi-interval measurement scales” (Ware and Gandek 1998). They argue that such scales can consistently rank health status, and that the ratio of differences between scores has meaning. While such claims can be debatable, it is important to note that all techniques used to convert discrete category scores into a continuous variable will have some issues. Evidence provided by Butterworth and Crosier (2004) supports the validity of SF-36 data collected by the HILDA survey as general measures of physical and mental health status. Scales for both physical and mental health are provided. We are concerned with the mental health scale. The mental health dimension of SF-36 consists of 5 multichoice questions that ask respondents about their perceptions of their mental health. The questions are:

1. “Been a very nervous person”
2. “Felt down in the dumps”

3. “Felt calm and peaceful”

4. “Felt downhearted and blue”

5. “Been a happy person”

Each respondent was asked to rate how often they felt in such a way in the past four weeks: [1] All the time, [2] Most of the time, [3] A good bit of the time, [4] Some of the time, [5] A little of time, [6] None of the time. The responses were scored using the scoring algorithm given in Ware et al. (1993). The final measure ranges between 0 and 100, where a score of 100 implies good mental health and a 0 represents a serious mental health problem. Individuals with scores below 50 are considered to have poor mental health.

Having obtained a mental health score for each individual in the sample, we are faced with the problem of using the sample of scores to estimate a distribution. If we adopt a parametric approach, then, given that each mental health score lies in the interval [0,100], a convenient continuous distribution for representing the population of scores is the beta distribution, applied to a scaling of the scores to make them lie in the interval [0,1]. Its probability density function (pdf) is given by

$$B(y \mid \alpha, \beta) = \frac{\Gamma(\alpha+\beta)}{\Gamma(\alpha)\Gamma(\beta)} y^{\alpha-1}(1-y)^{\beta-1}$$

where $y$ is a random mental health draw and $\alpha$ and $\beta$ are parameters. However, using a parametric distribution with only two parameters is unlikely to be adequate to capture a wide variety of mental health distributions. As an alternative, we use a Bayesian nonparametric approach, modelling an infinite mixture of beta distributions via a Dirichlet process prior (Escobar and West, 1995). In this context, the pdf for $y$ is given by

$$p(y \mid \boldsymbol{\alpha}, \boldsymbol{\beta}, \boldsymbol{w}) = \sum_{k=1}^{\infty} w_k B(y \mid \alpha_k, \beta_k) \tag{1}$$

with parameter vectors $\boldsymbol{\alpha}' = (\alpha_1, \alpha_2, \ldots)$, $\boldsymbol{\beta}' = (\beta_1, \beta_2, \ldots)$ and $\boldsymbol{w}' = (w_1, w_2, \ldots)$.

Distributions are estimated for each of the five years and for each of the population subgroups in those years, treating each set of observations as a cross-section, estimated with cross-sectional weights. Details of an MCMC algorithm for estimating this model are provided in the Supplementary Appendix. Draws from the posterior density of the parameters are taken at each iteration of the MCMC algorithm. Functions of these parameter draws are then used to estimate criteria for comparing distributions. The apparent need to sample an infinite number of parameters in (1) is avoided by using a device known as the slice sampler (Walker, 2007). At each of $j = 1, 2, \ldots, M$ iterations of the MCMC algorithm, this device stochastically truncates the infinite number of components in (1) to a finite number $K^{(j)}$. Posterior sampling for $\boldsymbol{\alpha}, \boldsymbol{\beta}$ and $\boldsymbol{w}$ proceeds using this finite number of elements; the draws $(\boldsymbol{\alpha}^{(j)}, \boldsymbol{\beta}^{(j)}, \boldsymbol{w}^{(j)})$ are of dimension $K^{(j)}$, with this dimension changing with each iteration. One "draw" from the predictive-posterior pdf for $y$ can be written as[3]

$$p^{(j)}\left(y \mid \boldsymbol{\alpha}^{(j)}, \boldsymbol{\beta}^{(j)}, \boldsymbol{w}^{(j)}\right) = \sum_{k=1}^{K^{(j)}} w_k^{(j)} B\left(y \mid \alpha_k^{(j)}, \beta_k^{(j)}\right) \tag{2}$$

The Bayesian nonparametric density estimate for $y$ is given by the average of (2) over all iterations. That is,

$$\hat{p}(y) = \frac{1}{M} \sum_{j=1}^{M} p^{(j)}\left(y \mid \boldsymbol{\alpha}^{(j)}, \boldsymbol{\beta}^{(j)}, \boldsymbol{w}^{(j)}\right) \tag{3}$$

For a given value of $y$, the spread of the values in (2) provides an indication of the reliability of (3) as an estimate of $p(y)$. The density in (3) is not used directly for comparing the health status of populations, but the quantities based on it are. Also, graphing (3), as we do in our empirical work, is useful for gaining an appreciation of the nature of the distribution. In the next section we describe the criteria we employ for health-status comparisons, with the exception of dominance, which we consider in Section 4.

## 3 Criteria for Comparing Distributions

Values of quantities used as criteria for comparing health status at different points in time and across different subgroups of population are computed from the MCMC output $(\boldsymbol{\alpha}^{(j)}, \boldsymbol{\beta}^{(j)}, \boldsymbol{w}^{(j)})$,

[3] In practice, a residual weight $w_{K+1}^{(j)} = 1 - \sum_{k=1}^{K^{(j)}} w_k^{(j)}$ is computed to complete the mixture and the sum in (2) and subsequent equations is a sum over $K^{(j)} + 1$ values for $w_k, \alpha_k$ and $\beta_k$. We have chosen to use $K^{(j)}$ in the summations in the body of the paper to avoid making an already complicated notation more complex. More precise details are provided in the Supplementary Appendix.

$j = 1, 2, \ldots, M$. The first of these is the mean health score; observations on it drawn from its posterior pdf are given by

$$\mu^{(j)} = \sum_{k=1}^{K^{(j)}} w_k^{(j)} m_k^{(j)} = \sum_{k=1}^{K^{(j)}} \frac{w_k^{(j)} \alpha_k^{(j)}}{\alpha_k^{(j)} + \beta_k^{(j)}} \tag{4}$$

where $m_k = \alpha_k / (\alpha_k + \beta_k)$ is the mean of the *k*-th component of the mixture. The average of the MCMC draws, $\hat{\mu} = \sum_{j=1}^{M} \mu^{(j)} / M$, is an estimate of the posterior mean which in turn is an estimate for $\mu$. The standard deviation of the $\mu^{(j)}$ is an estimate of the posterior standard deviation of $\mu$ and is an indication of the reliability of the estimate $\hat{\mu}$.

In a similar way, we can use averages and standard deviations of MCMC draws to estimate posterior means and standard deviations for other quantities of interest. In line with the MCMC algorithm, infinite sums in these quantities are truncated to $K^{(j)}$ in the *j*-th iteration of the algorithm. For comparing health status for those with acute mental illness, we consider properties of the distribution below a pre-specified threshold $z$. The proportion of the population below this threshold, known as the headcount ratio in the poverty literature, is given by

$$\begin{aligned} HC &= \int_0^z p(y \mid \boldsymbol{\alpha}, \boldsymbol{\beta}, \boldsymbol{w})\, dy \\ &= F(z \mid \boldsymbol{\alpha}, \boldsymbol{\beta}, \boldsymbol{w}) \\ &= \sum_{k=1}^{\infty} w_k F(z \mid \alpha_k, \beta_k) \end{aligned} \tag{5}$$

where $F(z \mid \boldsymbol{\alpha}, \boldsymbol{\beta}, \boldsymbol{w})$ is the cdf for $y$ and $F(y \mid \alpha_k, \beta_k) = \int_0^y B(t \mid \alpha_k, \beta_k) dt$ is the cdf of the *k*-th component.

A family of measures that considers not just the number of individuals below a threshold, but also the severity of the mental illness for those below the threshold, is that attributable to Foster et al. (1984). It is given by

$$FGT_a = \int_0^z \left( \frac{z - y}{z} \right)^a p(y \mid \boldsymbol{\alpha}, \boldsymbol{\beta}, \boldsymbol{w})\, dy$$

In our empirical work, we compare $FGT$ values for $a = 1$ and $a = 2$. For $a = 1$, we have

$$FGT_1 = HC - \frac{1}{z}\int_0^z y\, p(y \mid \boldsymbol{\alpha}, \boldsymbol{\beta}, \boldsymbol{w})dy$$

To write this quantity in a form convenient for calculation, we introduce the first-moment distribution for $y$ which is given by

$$F^{(1)}(y \mid \boldsymbol{\alpha}, \boldsymbol{\beta}, \boldsymbol{w}) = \frac{1}{\mu}\int_0^y t\, p(t \mid \boldsymbol{\alpha}, \boldsymbol{\beta}, \boldsymbol{w})dt$$

$$= \frac{1}{\mu}\sum_{k=1}^{\infty} w_k m_k F(y \mid \alpha_k + 1, \beta_k) \tag{6}$$

The second line of equation (6) is a convenient one for calculations. It shows that it is possible to write the first-moment distribution function with component parameters $\alpha_k$ and $\beta_k$ in terms of component cdf's whose parameters are $(\alpha_k + 1)$ and $\beta_k$. Using (6), $FGT_1$ becomes

$$FGT_1 = HC - \frac{\mu}{z} F^{(1)}(z \mid \boldsymbol{\alpha}, \boldsymbol{\beta}, \boldsymbol{w}) \tag{7}$$

In a similar way $FGT_2$ can be written in the more convenient form

$$FGT_2 = HC - \frac{2\mu}{z} F^{(1)}(z \mid \boldsymbol{\alpha}, \boldsymbol{\beta}, \boldsymbol{w}) + \frac{\mu_2}{z^2} F^{(2)}(z \mid \boldsymbol{\alpha}, \boldsymbol{\beta}, \boldsymbol{w}) \tag{8}$$

where $\mu_2 = \sum_{k=1}^{\infty} w_k \mu_{2,k}$ is the second moment for $y$, with $k$-th component $\mu_{2,k} = \alpha_k(\alpha_k + 1)/(\alpha_k + \beta_k)(\alpha_k + \beta_k + 1)$, and $F^{(2)}(z \mid \boldsymbol{\alpha}, \boldsymbol{\beta}, \boldsymbol{w})$ is the second-moment distribution function evaluated at $z$. It can be written as

$$F^{(2)}(z \mid \boldsymbol{\alpha}, \boldsymbol{\beta}, \boldsymbol{w}) = \frac{1}{\mu_2}\int_0^z y^2 p(y \mid \boldsymbol{\alpha}, \boldsymbol{\beta}, \boldsymbol{w})dy$$

$$= \frac{1}{\mu_2}\sum_{k=1}^{\infty} w_k \mu_{2,k} F(y \mid \alpha_k + 2, \beta_k) \tag{9}$$

The second line of (9) shows that the second-moment distribution function can be written in terms of component cdf's with parameters $\alpha_k + 2$ and $\beta_k$.

## 4. Assessing Stochastic Dominance

There are several ways that one can approach the question: has mental health of a population improved over time? Or, alternatively: is the mental health of one population subgroup better or worse than another? A simple way is to compare measures of central tendency such as the mean, median or mode, enabling one to say the “average” or “typical” level of mental health is better for one case than another. Using the posterior distribution for mean mental health, introduced in the previous section, is one example of this approach. Another more complete and more exacting way is to use a metric that compares whole distributions of mental health scores. Stochastic dominance concepts are useful for making such comparisons. Suppose we are comparing two mental health distributions that we call *A* and *B*, with corresponding beta-mixture cdfs, $F_A\left(y\,|\,\boldsymbol{\alpha}_A,\boldsymbol{\beta}_A,\boldsymbol{w}_A\right)$ and $F_B\left(y\,|\,\boldsymbol{\alpha}_B,\boldsymbol{\beta}_B,\boldsymbol{w}_B\right)$. We say that *A* first-order stochastically dominates *B* if and only if

$$F_A\left(y\,|\,\boldsymbol{\alpha}_A,\boldsymbol{\beta}_A,\boldsymbol{w}_A\right)\le F_B\left(y\,|\,\boldsymbol{\alpha}_B,\boldsymbol{\beta}_B,\boldsymbol{w}_B\right) \tag{10}$$

for all mental health scores $0\le y\le 1,$ with strict inequality holding for some $0<y<1.$ Using a simplified description of (10) that ignores equalities and the zero-one end points, $A_{FSD}B$ implies that proportion of people with a health score less than some value $y$ is less in population $A$ than it is in population $B;$ and this is true for all $0<y<1.$ Except at the endpoints, the cdf for $A$ lies to the right of the cdf for $B.$

We say that $A$ second order stochastically dominates $B$ if and only if

$$\int_0^y F_A\left(t\,|\,\boldsymbol{\alpha}_A,\boldsymbol{\beta}_A,\boldsymbol{w}_A\right)dt\le \int_0^y F_B\left(y\,|\,\boldsymbol{\alpha}_B,\boldsymbol{\beta}_B,\boldsymbol{w}_B\right)dt \tag{11}$$

for all $0\le y\le 1,$ with strict inequality holding for some $0<y<1.$ An intuitive interpretation of (11) is that the total mental health score for everybody with a score less than or equal to any mental health score $y$ is less in population $A$ than it is in population $B$. This condition is less strict than FSD; $A_{FSD}B$ implies $A_{SSD}B.$ It is possible for SSD to exist when the two cdf’s cross.

For our analysis of SSD, it turns out that there is a more convenient expression than that in (11). To derive this expression we use the first-moment distribution function given in equation (6). Using integration by parts, we can show that

$$\begin{aligned}\mu F^{(1)}\left(y\,|\,\boldsymbol{\alpha},\boldsymbol{\beta},\boldsymbol{w}\right) &= \int_0^y t\,p\left(t\,|\,\boldsymbol{\alpha},\boldsymbol{\beta},\boldsymbol{w}\right)dt \\ &= yF\left(y\,|\,\boldsymbol{\alpha},\boldsymbol{\beta},\boldsymbol{w}\right)-\int_0^y F\left(t\,|\,\boldsymbol{\alpha},\boldsymbol{\beta},\boldsymbol{w}\right)dt\end{aligned}$$

Hence, $A_{SSD}B$ if and only if

$$yF_A\left(y\,|\,\boldsymbol{\alpha}_A,\boldsymbol{\beta}_A,\boldsymbol{w}_A\right)-\mu_A F_A^{(1)}\left(y\,|\,\boldsymbol{\alpha}_A,\boldsymbol{\beta}_A,\boldsymbol{w}_A\right)\le yF_B\left(y\,|\,\boldsymbol{\alpha}_B,\boldsymbol{\beta}_B,\boldsymbol{w}_B\right)-\mu_B F_B^{(1)}\left(y\,|\,\boldsymbol{\alpha}_B,\boldsymbol{\beta}_B,\boldsymbol{w}_B\right) \quad (12)$$

for all $0\le y\le 1$ and with strict inequality holding for at least some $0<y<1.$ Equation (12) is the one used to compare posterior probabilities of dominance, along the lines that we now discuss.

To assess whether conditions (10) and/or (12) hold for two specified populations, the distribution functions $F,$ and the first moment distribution functions $F^{(1)}$ need to be estimated using samples from each of the populations. Then, recognizing the existence of sampling error, we need a way of presenting the degree to which the sample information supports the existence of dominance in either direction, or the existence of no dominance. To make this inference problem specific, let

$$D_1(y)=F_B(y)-F_A(y) \quad (13)$$

and

$$D_2(y)=\left[yF_B(y)-\mu_B F_B^{(1)}(y)\right]-\left[yF_A(y)-\mu_A F_A^{(1)}(y)\right] \quad (14)$$

where $\left[F_A(y),F_A^{(1)}(y),F_B(y),F_B^{(1)}(y)\right]$ is abbreviated notation for the functions in (10) and (12). Let $\hat{D}_1(y)$ and $\hat{D}_2(y)$ denote estimates of the functions $D_1(y)$ and $D_2(y),$ obtained by replacing $\left[\mu_A,F_A(y),F_A^{(1)}(y),\mu_B,F_B(y),F_B^{(1)}(y)\right]$ by their estimates, $\left[\hat{\mu}_A,\hat{F}_A(y),\hat{F}_A^{(1)}(y),\hat{\mu}_B,\hat{F}_B(y),\hat{F}_B^{(1)}(y)\right].$ For $A$ to dominate $B$ we require $D_i(y)\ge 0$ for all $0\le y\le 1$ and $D_i(y)>0$ for some $0<y<1,$ where $i=1$ for FSD, and $i=2$ for SSD. For $B$ to dominate $A,$ the inequalities for $D_i(y)$ are reversed.

A vast number of sampling theory tests that typically use a nonparametric estimate $\hat{D}_i(y)$ to test for dominance have appeared in the literature[4]. Special problems arise because $D_i(y)$ is a function and not a point. Some tests resolve this issue by estimating $D_i(y)$ at a number of points and using its maximum value as a test statistic. Others base a test statistic on the joint distribution of $D_i(y)$ evaluated at a number of points. Also relevant is whether dominance or no dominance is chosen as the null hypothesis; this choice has a bearing on what conclusions are possible. As described by Davidson and

[4] This literature can be accessed through the extensive list of references in Lander et al (2020).

Duclos (2013), if dominance is specified as the null hypothesis, dominance of one distribution over another can never be established. Failure to reject a null that $A$ dominates $B$ can mean that (a) $A$ dominates $B$, (b) neither distribution is dominant, or (c) that $A$ does not dominate $B$, but the magnitude of the test statistic was not sufficiently large to conclude it was "significant". Reversal of the distributions so that $B$ dominates $A$ is the null hypothesis does little to resolve this dilemma. If both $A$ dominating $B$ and $B$ dominating $A$ are rejected, one can conclude with some confidence that neither distribution is dominant. However, other combinations of outcomes do not lead to a firm conclusion. Changing the null hypothesis to, say, $A$ does not dominate $B$ has the advantage that it leads to a legitimate claim that $A$ dominates $B$ if the null hypothesis is rejected. However, as pointed out by Davidson and Duclos (2013), for continuous distributions such a null hypothesis can never be rejected unless the range of the variable being considered is restricted. The problem lies in the tails where the distributions converge. In our case, for FSD, we have $D_1(0) = D_1(1) = 0,$ and, for SSD, we have $D_2(0) = 0$ and $D_2(1) = \mu_A - \mu_B$ [5].

In contrast to the sampling theory approach, we do not use a formal hypothesis testing framework for summarizing the sample information about dominance. Instead, following earlier work (Lander et al 2020, Gunawan et al 2020), we avoid the need to specify a null hypothesis by computing the posterior probabilities for each possible outcome: $A$ dominates $B$, $B$ dominates $A$, and neither distribution is dominant. To compute these probabilities we begin with the MCMC draws for $\boldsymbol{\theta} = \left(\boldsymbol{\alpha}_A, \boldsymbol{\beta}_A, \boldsymbol{w}_A, \boldsymbol{\alpha}_B, \boldsymbol{\beta}_B, \boldsymbol{w}_B\right)$ from their posterior distributions. For each value of $y$ in the interval [0,1], there will be posterior distributions for $D_1(y)$ and $D_2(y)$ implied by the posterior distributions for the elements in $\boldsymbol{\theta}$. For each MCMC draw for $\boldsymbol{\theta}$ there are corresponding draws for $D_1(y)$ and $D_2(y)$ for every value of $y$. Let $D_i^{(j)}(y),\ i = 1,2$, be the *j*-th MCMC draw on $D_i(y)$. Then, an estimate of the probability that $D_i(y)$ is non-negative at the point $y$ is equal to the proportion of draws for which $D_i^{(j)}(y) \geq 0.$ That is,

---

[5] This last result shows that $\mu_A \geq \mu_B$ is a necessary condition for SSD.

$$\Pr\left[D_i(y) \geq 0\right] = \frac{1}{M}\sum_{j=1}^{M} I\left[D_i^{(j)}(y) \geq 0\right] \tag{15}$$

where $I[\cdot]$ is an indicator function equal to one if its argument is true and zero otherwise. To evaluate the probability of dominance, we need the proportion of draws for which $D_i^{(j)}(y) \geq 0$ for <u>all</u> $y,$ and with $D_i^{(j)}(y) > 0$ for at least one $y.$ Because $y$ is continuous, the best we can do in terms of evaluating $D_i^{(j)}(y)$ at all values of $y,$ is to compute it for a fine grid of values in the interval $0 \leq y \leq 1.$ Except at, or close to, the endpoints, it is unlikely that computed values of $D_i^{(j)}(y)$ will be exactly zero, and so, whether we treat $D_i^{(j)}(y) \geq 0$ as a strict or non-strict inequality is immaterial. Given this framework, estimates of the posterior probabilities can be specified as

$$\Pr(A \text{ dominates } B) = \frac{1}{M}\sum_{j=1}^{M}\prod_{h=1}^{H} I\left[D_i^{(j)}(y_h) \geq 0\right]$$

$$\Pr(B \text{ dominates } A) = \frac{1}{M}\sum_{j=1}^{M}\prod_{h=1}^{H} I\left[D_i^{(j)}(y_h) \leq 0\right]$$

$$\Pr(\text{neither distribution dominates}) = 1 - \Pr(A \text{ dominates } B) - \Pr(B \text{ dominates } A)$$

where $y_1, y_2, \ldots, y_H$ is the selected grid of values of $y$ in the interval [0,1]. As one might expect, the results can be sensitive to the endpoints selected for the grid of $y$ values. At 0 and 1 for FSD and 0 for SSD, the variance of the posterior distributions collapses to zero. Close to these points there will be only a small variation which can impact on the dominance probabilities. This impact can be monitored by using equation (15) to plot $\Pr\left[D_i(y) \geq 0\right]$ against $y,$ yielding curves that we call “probability curves”. Because

$$\frac{1}{M}\sum_{j=1}^{M}\prod_{h=1}^{H} I\left[D_i^{(j)}(y_h) \geq 0\right] \leq \min_{y_h}\left\{\frac{1}{M}\sum_{j=1}^{M} I\left[D_i^{(j)}(y_h) \geq 0\right]\right\}$$

these curves provide a convenient device for checking the values of $y$ having the greatest effect on the probability of dominance. Excluding values of $y$ close to the endpoints is akin to Davidson and Duclos (2013) notion of restricted dominance. It may be less than ideal, but we can always document the range being considered, choose ranges that are of particular interest, and examine sensitivity to changes in the range.

There are two main differences in the methodology used here to assess mental health distributions and that previously used to examine dominance of income distributions. The earlier studies used a mixture of gamma densities to model income distributions, a finite mixture in Lander et al (2020) and an infinite mixture in Gunawan et al (2020). The other difference is in the conditions used to evaluate FSD and SSD. To avoid subjectively specifying a maximum income for the grid of income values, the earlier studies expressed the conditions for FSD and SSD in terms of quantile functions and checked for dominance over a range of population proportions that must lie between zero and one. For our mental health distributions where the mental health score lies in the [0,1] interval, it is natural to express the FSD and SSD conditions in terms of their distribution functions and compute dominance probabilities along these lines.

## 5. Results

We first examine changes in the mental health distribution over time for the complete population, and then compare distributions for the sub-populations male/female and indigenous/non indigenous.

### 5.1 Comparing Distributions over Time

Estimates of the health densities, obtained by averaging the functions at each of the MCMC draws of the parameters (see equation (3)), are plotted in Figure 1 for the years 2001, 2006, 2010, 2014 and 2017. They are negatively skewed and bimodal, with one mode lying between 0.8 and 0.9 and a secondary mode at the maximum value of 0.999.[6] There is very little difference between the distributions; those for 2014 and 2017 are slightly to the left of those for 2006 and 2010. The sample means, reported in Table 1 along with other summary statistics, are consistent with this observation. Those for 2014 and 2017 are slightly less than those for 2006 and 2010. The posterior densities for mean health in each of the years are graphed in Figure 2 with the means and standard deviations of these densities reported in Table 2. They suggest mental health improved from 2001 to 2006/2010, but declined thereafter.

[6] Health scores at 1.000 were set to 0.999 to avoid instability in the estimation process. The percentages of observations changed from 1.000 to 0.999 were 3.2% in 2001, 2.7% in 2006, 2.3% in 2010, 2.1% in 2014, and 2.5% in 2017.

To assess whether there are fewer people with poor mental health in the later years, and whether the severity of mental health has lessened over time, we examine the headcount ratio and the FGT indices with poor mental health defined as those scores less than 0.5. Their values, calculated from the raw data, are presented in Table 1, and their estimates (posterior means) from the model are presented in Table 2. All these measures decrease from 2001 to 2010 and increase thereafter suggesting an initial improvement in the severity of mental health but a decline in the later years.

A more complete picture of whether or not there has been an improvement in the mental health distributions is obtained by comparing distribution functions, and computing their dominance probabilities. From the distributions functions plotted in Figure 3, there is no clear ranking between all the years. The FSD probabilities reported in Table 3 support this observation. They were computed using 1000 equal spaced values for the health score from 0.1 to 0.999. The left-tail cut off of 0.1 was considered reasonable given the small number of observations below this value. Except for 2017 versus 2006 where the probability of no dominance is 0.69, all probabilities for no dominance are greater than 0.84, and most are greater than 0.9.

To gain more insights into these results, it is useful to examine a sample of the probability curves. When examining just the mean scores, we find 2006 and 2010 have the highest score – they are approximately the same – and 2017 has the lowest. The probabilities for the years with the highest mean scores dominating that with the lowest mean score are $\Pr(2006_{FSD}2017) = 0.308$ and $\Pr(2010_{FSD}2017) = 0.062.$ The behaviour of the probability curves, depicted in Figure 4, shows why these two probabilities differ. Both curves suggest that, for scores in the interval $0.3 \le y \le 0.7,$ the probability of dominance would be close to 1. Including the tails leads to a reduction in these probabilities, with the reduction attributable to the right tail being particularly dramatic for $2010_{FSD}2017.$ Also included in Figure 4, with scale on the right axis, are the posterior means for $D_1^{(a)}(y) = F_{2017}(y) - F_{2006}(y)$ and $D_1^{(b)}(y) = F_{2017}(y) - F_{2010}(y).$ As expected, the larger values for $\Pr\left[D_1(y) \ge 0\right]$ correspond to the largest values for $F_{2017}(y) - F_{2006}(y).$ In Figure 4(b) where the curves for $2010_{FSD}2017$ are plotted, the estimated cdf's cross leading to some negative values for $D_1(y)$ and a very low probability of dominance.

When we move to consider the less strict criterion of SSD, we find more evidence of dominance of some years over the others. From Table 4, the probabilities for 2001 being dominated by 2006 and 2010 are 0.548 and 0.809, respectively; these represent large increases from the FSD values of 0.036 and 0.001. Comparing 2006 and 2010 with 2017, we find the probability that 2006 dominates 2017 has increased from 0.308 to 0.674, while that for 2010 over 2017 has increased from 0.062 to 0.799. These increases can be attributed to behaviour in the upper tails. As illustrated in Figure 5, the probability curves remain at one, in the upper tail, and the values for $D_2(y)$ remain large, unlike those for FSD.

Summarizing the FSD and SSD information, we conclude that there is little evidence to suggest that complete health-score distribution has shifted to the right or left over time, making every segment of the population better or worse off. However, in terms of the accumulated health score for every segment of the population, there is relatively strong evidence to suggest improvement from 2001 to 2010 and a decline thereafter.

Also of interest are the dominance results for those with poor mental health. The single index measures of headcount, $FGT_1$ and $FGT_2$ reported in Tables 1 and 2 for scores less than 0.5 suggest an improvement from 2001 to 2010 and a decline thereafter, results that are in line with the SSD results for the whole population. The dominance results in Tables 5 and 6 support this conclusion. We have $\Pr\left(2010_{FSD}2001\right) = 0.805$, $\Pr\left(2010_{SSD}2001\right) = 0.838$, $\Pr\left(2010_{FSD}2017\right) = 0.795$, and $\Pr\left(2010_{SSD}2017\right) = 0.800$. There is less evidence of dominance when more adjacent years are considered; 2010 dominating 2014 is an exception where both probabilities are greater than 0.7.

### 5.2 Male and Female Subgroups

Dominance criteria can also be used to track the welfare over time for subgroups of the population and to compare subgroups at a particular point in time. In this section we consider the male/female subgroups. Indigenous/non-indigenous subgroups are considered in the next section. Table 7 contains the sample statistics for the male/female subgroups in each year. Examining mean health suggests male mental health scores are substantially above female mental health scores, and that changes in mental health scores over time are consistent with those for the complete sample. The headcount, $FGT_1$ and $FGT_2$ indices for males are smaller than for females for all the years being considered.

When we move from a male/female comparison based on means to one based on their complete distributions we find moderate evidence that the male distribution dominates the female distribution with respect to FSD, and relatively strong evidence of dominance with respect to SSD. The probabilities are reported in Table 8: $\Pr\left(\text{Male}_{FSD}\text{Female}\right)$ falls from 0.428 in 2001 to 0.295 in 2006, but then increases to 0.705 in 2017. The balance of the probability is for no dominance. In all years, there is a zero probability that the female distribution dominates the male distribution. The SSD results tell a similar story, but with higher probabilities. The probabilities for dominance by the male distribution decrease from 0.540 in 2001 to 0.418 in 2006, and then increase to 0.914 in 2017. In this context, males are healthier than females and the gap has been widening over the decade from 2006 to 2017. If both tails of the distributions are ignored, then the probabilities for male being dominant, for both FSD and SSD, are close to 1 in every year. We illustrate this fact by plotting 2017 FSD and SSD probability curves in Figure 6. When the ranges for these two curves are restricted to $0.4 < y < 0.85$ for FSD and to $0.4 < y < 0.999$ for SSD, the dominance probabilities for FSD and SSD are close to 1.

Our earlier analysis of the whole population suggested mental health has deteriorated from 2010 to 2017. When we combine this information with evidence that female mental health is worse than male mental health, and that the probability of it being worse increased from 2006 to 2017, we would expect the dominance probabilities of earlier years over later years to be greater for females than for males. This expectation is realized for SSD but not for FSD. The complete set of FSD and SSD probabilities for male and female subgroups for all years can be found in Table S2-S5 in the Supplementary Appendix. The FSD probabilities over time for both subgroups show little evidence of dominance in either direction; the probabilities for no dominance are all greater than 0.84, with most greater than 0.9. The SSD probabilities, reported in Table 9 for dominance of each of the earlier years over each of the later years for the years 2006 onwards, are more in line with our expectations. Except for $2010_{SSD}2014$, the probabilities of a deteriorating health distribution are all greater for females than they are for males.

Considering a female-male comparison for only those with poor health – those with a score below 0.5 – leads to similar conclusions to those reached for all males and females. Increases in the probability of dominance are expected since the number of draws satisfying dominance in the restricted range must be at least as great as the number over the complete range. From Table 10, we note that, relative to Table 8, the probabilities for the male distribution dominating the female distribution are only slightly greater for SSD, but they are considerable greater for FSD. For example, in 2014 we have $\Pr\left(\text{Male}_{FSD}\text{Female}\right) = 0.560$ and $\Pr\left(\text{Male}_{FSD}\text{Female} \mid y < 0.5\right) = 0.813$; in 2017 the values are $\Pr\left(\text{Male}_{FSD}\text{Female}\right) = 0.705$ and $\Pr\left(\text{Male}_{FSD}\text{Female} \mid y < 0.5\right) = 0.896.$ The larger changes for FSD can be attributed to removal of the right tail of the distribution. From Figure 6 we observe that removing the right tail has little impact on the SSD probability, but it increases the probability for FSD. Overall, we conclude that the incidence of poor mental health is greater for females and the disparity has increased from 2006 to 2017. Also, from results probabilities in Table 11, we see that the probabilities for dominance of later years over earlier years are greater for males than females when mental health is improving (2001 to 2010), and the probabilities for dominance of the earlier years over the later years is greater for females than males when mental health is deteriorating (2010 to 2017). In other words, the evidence suggests the mental health of males improves more than that of females when mental health of the whole population is improving, and it deteriorates less than that of females when mental health is deteriorating. The full set of probabilities is given in Tables S6 to S9 in the Supplementary Appendix.

### 5.3 Indigenous and Non-indigenous Subgroups

In this section we compare the mental health distributions of indigenous and non-indigenous subgroups in each of the years 2001, 2006, 2010, 2014 and 2017, as well as how the distributions of each subgroup have changed over time. We have excluded Torres Strait Islanders and focused on the Aboriginal/Non-aboriginal subgroups. A major policy objective of the federal government is to narrow the health gap between the aboriginal population and the remainder of Australians. We examine the evidence for the narrowing of this gap in the context of mental health distributions. Table 12 contains the sample means, standard deviations, sample headcount, $FGT_1$ and $FGT_2$ indices for each subgroup and year.[7] The mean mental health scores for the non-indigenous sample are substantially above those for the indigenous sample. Also, the gap between the two is relatively constant between 2001 and 2010, but increases dramatically in 2014, and then decreases slightly in 2017, leaving a gap larger in 2017 than it was in 2001. Considering the proportion of people with poor mental health (the headcount), and the severity of the poor mental health $(FGT_1$ and $FGT_2),$ reveals a similar story. From 2001 to 2017, the proportion of people with poor mental health and its severity increased for both groups, and the gap between indigenous and non-indigenous widened. The change was not monotonic, however. For the indigenous sample, there was a deterioration from 2001 to 2006, an improvement from 2006 to 2010, a large deterioration from 2010 to 2014, and then an improvement from 2014 to 2017.

The much smaller sample sizes for the indigenous population – about 2-3% of the total sample size – means that their population quantities are estimated less accurately than those for the non-indigenous population. To illustrate the difference, in Figures 7 to 8 we plot the posterior densities for the means and the headcounts for the years 2001 and 2017. The greater uncertainty associated with the indigenous measures is reflected in wider dispersion in their posterior distributions, but there is still ample evidence to support a deterioration in mental health and a widening of the gap.

Dominance probabilities for comparing the indigenous/non-indigenous subgroups are reported in Table 13 for their complete populations $(y < 1)$ and for those with poor mental health $(y \leq 0.5).$ For the complete populations and FSD, no dominance is the most likely outcome with the probabilities for no dominance all being greater than 0.82 in all years except in 2014 where the no-dominance probability was 0.62. This result is perhaps surprising given the differences in the means, differences that are greater than those from the male/female comparison which yielded greater probabilities for dominance. It can be explained by the smaller sample size for the indigenous subgroup. This smaller sample size results in greater dispersion of $D_1(y)$ and $D_2(y)$ at each point $y,$ increasing the probability that their MCMC draws can be both positive and negative. The highest probability for first-order dominance of the indigenous distribution by the non-indigenous distribution is 0.38 in 2014, which is the year with the largest difference in mean health scores. When we consider FSD for those with poor mental health, there is more evidence of dominance of the indigenous group by the non-indigenous group. It is the most likely outcome in all years except 2001 with the highest probability being 0.94 in 2014. With SSD, the results suggest dominance of the indigenous group by the non-indigenous group for both the complete population and those with poor mental health. The two sets of probabilities are almost identical, with the 2014 probability being 0.95 and the probabilities in the other years ranging between 0.42 and 0.71. For both FSD and SSD, the probability of the indigenous distribution dominating the non-indigenous distribution is close to zero. In terms of closing the gap between the two distributions, there is a narrowing of the gap between 2014 and 2017, but the gap in 2017 remains wider than it was in 2001. In Figure 9 we plot the probability curves and $D_1(y)$ and $D_2(y)$ for the non-indigenous distribution dominating the indigenous distribution in 2001 and 2017. It is clear why the FSD probabilities for non-indigenous over indigenous are so much less than those for SSD. There is a sharp decline in the FSD probability curves for $y > 0.9,$ with the mean cdf's crossing at approximately $y = 0.95.$ In contrast, the SSD probability curves reach their highest point when $y > 0.8.$ The sharp decline in the FSD probability curves also explains why the FSD dominance probabilities are considerably higher when we consider only those with poor mental health. Finally, the SSD probability curves having their highest values for large $y$ explains why the SSD dominance probabilities for the whole populations are approximately equal to those for the population where $y \leq 0.5.$

[7] The posterior means and standard deviations for a selection of these quantities are given in Table S10 of the Supplementary Appendix. They match closely those from the raw data.

Dominance probabilities reflecting how the indigenous distribution has changed over time are reported in Table 14. It is difficult to establish FSD of one year over any other year; the probabilities for no dominance are all greater than 0.72. When we move to SSD, there is stronger evidence of a mental health distribution that deteriorates over time. There are smaller probabilities of improvement from 2006 to 2010 and from 2014 to 2017 (0.34 and 0.44, respectively), but the probability that 2001 is preferred to 2006 is 0.38, and the probability that 2010 is preferable to 2014 is 0.73. Overall, from 2001 to 2017 there is a 0.52 probability that 2001 dominates 2017, and a probability of 0.44 that there is no dominance. When we consider the subset of the population with poor mental health - scores below 0.5 – there is stronger evidence for improvement in the periods 2006–2010 and 2014–2017, and stronger evidence for deterioration in the other intervals 2001–2006 and 2010–2014. Overall, the FSD and SSD probabilities for worsening of poor mental health over the complete period 2001–2017 are 0.5 and 0.55, respectively. A complete set of probabilities is given in Tables S11 to S14 of the Supplementary Appendix.

## 5. Conclusions

We have used stochastic dominance concepts to present evidence on how the Australian mental health distribution has changed over time and to compare distributions for male/female and indigenous/non-indigenous subgroups of the population. Summarising the evidence leads to the following key findings:

1) In terms of SSD, there were improvements in mental health from 2001 to 2010, but a deterioration from 2010 to 2017. This conclusion holds for the whole population, for both males and females, and for those with poor mental health.
2) When considering the stricter criterion of FSD, the conclusions in (1) also hold for the subset of the population with poor mental health. However, it is difficult to establish dominance in either direction for the whole population.
3) Males experience a better level of mental health than females, and this is particularly the case at levels of mental health considered poor. There is also evidence that the gap between female and male mental health has been widening towards the end of the sample.
4) The relatively small sample from the indigenous population makes conclusions about the indigenous/non-indigenous comparison less definite, but the results point towards poorer mental health for the indigenous population and a widening of the gap between indigenous and non-indigenous mental health.

Drawing these conclusions using the criterion of stochastic dominance is a novel and more exacting approach that considers complete distributions of mental health rather than single statistics such as the mean. The conclusions do have to be qualified, however, because they do depend on the validity of the scores as a representation of mental health, and on the legitimacy of comparing distributions over time and for different population subgroups. We need to assume similar responses to the questionnaire in different time periods, or from different population subgroups, are indicative of the same level of mental health.

**Table 1: Summary Statistics for Health Scores**

| | 2001 | 2006 | 2010 | 2014 | 2017 |
|---|---|---|---|---|---|
| Sample Mean | 0.7351 | 0.7407 | 0.7410 | 0.7373 | 0.7291 |
| Minimum | 0.0400 | 0.0400 | 0.0400 | 0.0400 | 0.0400 |
| Maximum | 0.9990 | 0.9990 | 0.9990 | 0.9990 | 0.9990 |
| Standard Deviation | 0.1748 | 0.1717 | 0.1679 | 0.1739 | 0.1789 |
| Sample Size | 12873 | 11545 | 11911 | 15387 | 15781 |
| Headcount (cutoff = 0.5) | 0.1049 | 0.0978 | 0.0949 | 0.1051 | 0.1147 |
| $FGT_1$ (cutoff = 0.5) | 0.0261 | 0.0235 | 0.0217 | 0.0262 | 0.0293 |
| $FGT_2$ (cutoff = 0.5) | 0.0111 | 0.0098 | 0.0086 | 0.0110 | 0.0125 |
| Proportion below 0.1 | 0.0021 | 0.0017 | 0.0013 | 0.0021 | 0.0020 |

**Table 2: Posterior Means (Standard Deviations) for Health Score Matrixs**

| | 2001 | 2006 | 2010 | 2014 | 2017 |
|---|---|---|---|---|---|
| Means | 0.7346<br>(0.0028) | 0.7406<br>(0.0029) | 0.7407<br>(0.0028) | 0.7368<br>(0.0028) | 07287<br>(0.0028) |
| Headcount | 0.1075<br>(0.0047) | 0.1002<br>(0.0049) | 0.0962<br>(0.0044) | 0.1066<br>(0.0042) | 0.1173<br>(0.0046) |
| $FGT_1$ | 0.0263<br>(0.0016) | 0.0236<br>(0.0017) | 0.0216<br>(0.0015) | 0.0259<br>(0.0015) | 0.0293<br>(0.0015) |
| $FGT_2$ | 0.0112<br>(0.0009) | 0.0099<br>(0.0009) | 0.0085<br>(0.0008) | 0.0109<br>(0.0008) | 0.0125<br>(0.0009) |

*Notes*: The values are those from the Beta mixture model. They match closely those calculated from the raw data, reported in Table 1.

**Table 3: First Order Stochastic Dominance Probabilities**

| | *A* | *B* | *A* | *B* | *A* | *B* | *A* | *B* |
|---|---|---|---|---|---|---|---|---|
| | 2017 | 2014 | 2014 | 2010 | 2010 | 2006 | 2006 | 2001 |
| Pr($A_{FSD}B$) | 0.0001 | | 0.0005 | | 0.0022 | | 0.0361 | |
| Pr($B_{FSD}A$) | 0.0256 | | 0.0874 | | 0.0094 | | 0.0003 | |
| Pr(no dominance) | 0.9743 | | 0.9121 | | 0.9884 | | 0.9636 | |
| | | | 2017 | 2010 | 2014 | 2006 | 2010 | 2001 |
| Pr($A_{FSD}B$) | | | 0.0000 | | 0.0004 | | 0.0011 | |
| Pr($B_{FSD}A$) | | | 0.0624 | | 0.1033 | | 0.0000 | |
| Pr(no dominance) | | | 0.9376 | | 0.8963 | | 0.9989 | |
| | | | | | 2017 | 2006 | 2014 | 2001 |
| Pr($A_{FSD}B$) | | | | | 0.0000 | | 0.0006 | |
| Pr($B_{FSD}A$) | | | | | 0.3083 | | 0.0041 | |
| Pr(no dominance) | | | | | 0.6917 | | 0.9953 | |
| | | | | | | | 2017 | 2001 |
| Pr($A_{FSD}B$) | | | | | | | 0.0001 | |
| Pr($B_{FSD}A$) | | | | | | | 0.1505 | |
| Pr(no dominance) | | | | | | | 0.8494 | |

**Table 4: Second Order Stochastic Dominance Probabilities**

| | *A* 2017 / *B* 2014 | *A* 2014 / *B* 2010 | *A* 2010 / *B* 2006 | *A* 2006 / *B* 2001 |
|---|---|---|---|---|
| $\Pr(A_{SSD} B)$ | 0.0024 | 0.0029 | 0.2805 | 0.5479 |
| $\Pr(B_{SSD} A)$ | 0.4797 | 0.6730 | 0.0501 | 0.0087 |
| Pr(no dominance) | 0.5179 | 0.3241 | 0.6694 | 0.4434 |
| | | 2017 / 2010 | 2014 / 2006 | 2010 / 2001 |
| $\Pr(A_{SSD} B)$ | | 0.0004 | 0.0302 | 0.8094 |
| $\Pr(B_{SSD} A)$ | | 0.7985 | 0.4093 | 0.0009 |
| Pr(no dominance) | | 0.2011 | 0.5605 | 0.1897 |
| | | | 2017 / 2006 | 2014 / 2001 |
| $\Pr(A_{SSD} B)$ | | | 0.0003 | 0.2971 |
| $\Pr(B_{SSD} A)$ | | | 0.6744 | 0.0559 |
| Pr(no dominance) | | | 0.3253 | 0.6470 |
| | | | | 2017 / 2001 |
| $\Pr(A_{SSD} B)$ | | | | 0.0084 |
| $\Pr(B_{SSD} A)$ | | | | 0.3740 |
| Pr(no dominance) | | | | 0.6176 |

**Table 5: First Order Stochastic Dominance Probabilities for the Population with a Mental Health Score Below 0.5**

| | *A* 2017 – *B* 2014 | *A* 2014 – *B* 2010 | *A* 2010 – *B* 2006 | *A* 2006 – *B* 2001 |
|---|---|---|---|---|
| Pr($A_{FSD} B$) | 0.0106 | 0.0021 | 0.3114 | 0.4746 |
| Pr($B_{FSD} A$) | 0.4482 | 0.7371 | 0.0332 | 0.0137 |
| Pr(no dominance) | 0.5412 | 0.2608 | 0.6554 | 0.5117 |
| | | 2017 – 2010 | 2014 – 2006 | 2010 – 2001 |
| Pr($A_{FSD} B$) | | 0.0003 | 0.0197 | 0.8052 |
| Pr($B_{FSD} A$) | | 0.7954 | 0.4214 | 0.0011 |
| Pr(no dominance) | | 0.2043 | 0.5589 | 0.1937 |
| | | | 2017 – 2006 | 2014 – 2001 |
| Pr($A_{FSD} B$) | | | 0.0003 | 0.2157 |
| Pr($B_{FSD} A$) | | | 0.6595 | 0.1129 |
| Pr(no dominance) | | | 0.3402 | 0.6714 |
| | | | | 2017 – 2001 |
| Pr($A_{FSD} B$) | | | | 0.0098 |
| Pr($B_{FSD} A$) | | | | 0.3609 |
| Pr(no dominance) | | | | 0.6293 |

**Table 6: Second Order Stochastic Dominance Probabilities for the Population with a Mental Health Score below 0.5**

| | *A* 2017 – *B* 2014 | *A* 2014 – *B* 2010 | *A* 2010 – *B* 2006 | *A* 2006 – *B* 2001 |
|---|---|---|---|---|
| Pr($A_{SSD}B$) | 0.0440 | 0.0058 | 0.4674 | 0.6029 |
| Pr($B_{SSD}A$) | 0.4821 | 0.7760 | 0.0767 | 0.0405 |
| Pr(no dominance) | 0.4739 | 0.2182 | 0.4559 | 0.3566 |
| | | 2017 – 2010 | 2014 – 2006 | 2010 – 2001 |
| Pr($A_{SSD}B$) | | 0.0009 | 0.0578 | 0.8382 |
| Pr($B_{SSD}A$) | | 0.7999 | 0.5170 | 0.0038 |
| Pr(no dominance) | | 0.1992 | 0.4252 | 0.1580 |
| | | | 2017 – 2006 | 2014 – 2001 |
| Pr($A_{SSD}B$) | | | 0.0022 | 0.3603 |
| Pr($B_{SSD}A$) | | | 0.6773 | 0.1787 |
| Pr(no dominance) | | | 0.3205 | 0.4610 |
| | | | | 2017 – 2001 |
| Pr($A_{SSD}B$) | | | | 0.0466 |
| Pr($B_{SSD}A$) | | | | 0.3900 |
| Pr(no dominance) | | | | 0.5634 |

**Table 7: Summary Statistics for the Health Scores**

| | 2001 | | 2006 | | 2010 | | 2014 | | 2017 | |
|---|---|---|---|---|---|---|---|---|---|---|
| | Male | Female | Male | Female | Male | Female | Male | Female | Male | Female |
| Sample Mean | 0.7465 | 0.7242 | 0.7530 | 0.7291 | 0.7533 | 0.7292 | 0.7475 | 0.7277 | 0.7399 | 0.7189 |
| Minimum | 0.0400 | 0.0400 | 0.0400 | 0.0400 | 0.0400 | 0.0400 | 0.0400 | 0.0400 | 0.0400 | 0.0400 |
| Maximum | 0.9990 | 0.9990 | 0.9990 | 0.9990 | 0.9990 | 0.9990 | 0.9990 | 0.9990 | 0.9990 | 0.9990 |
| Standard Deviation | 0.1719 | 0.1768 | 0.1670 | 0.1752 | 0.1619 | 0.1726 | 0.1688 | 0.1780 | 0.1731 | 0.1836 |
| Sample Size | 6067 | 6806 | 5379 | 6166 | 5587 | 6324 | 7214 | 8173 | 7405 | 8376 |
| Headcount (cutoff = 0.5) | 0.0978 | 0.1116 | 0.0806 | 0.1140 | 0.0773 | 0.1118 | 0.0943 | 0.1154 | 0.0996 | 0.1291 |
| $FGT_1$ (cutoff = 0.5) | 0.0237 | 0.0283 | 0.0200 | 0.0268 | 0.0170 | 0.0261 | 0.0221 | 0.0300 | 0.0247 | 0.0337 |
| $FGT_2$ (cutoff = 0.5) | 0.0100 | 0.0121 | 0.0084 | 0.0111 | 0.0067 | 0.0103 | 0.0090 | 0.0129 | 0.0102 | 0.0146 |
| Proportion below 0.1 | 0.0016 | 0.0025 | 0.0018 | 0.0016 | 0.0006 | 0.0013 | 0.0013 | 0.0028 | 0.0012 | 0.0029 |

**Table 8: Dominance probabilities for Male VS Female Subgroups**

| | 2001 | 2006 | 2010 | 2014 | 2017 |
|---|---|---|---|---|---|
| Pr(Female$_{FSD}$Male) | 0.0000 | 0.0000 | 0.0000 | 0.0000 | 0.0000 |
| Pr(Male$_{FSD}$Female) | 0.4278 | 0.2947 | 0.4870 | 0.5603 | 0.7045 |
| Pr(No FSD) | 0.5722 | 0.7053 | 0.5130 | 0.4397 | 0.2955 |
| Pr(Female$_{SSD}$Male) | 0.0000 | 0.0000 | 0.0001 | 0.0001 | 0.0000 |
| Pr(Male$_{SSD}$Female) | 0.5404 | 0.4176 | 0.6624 | 0.9139 | 0.9139 |
| Pr(No SSD) | 0.4596 | 0.5824 | 0.3375 | 0.0861 | 0.0861 |

**Table 9: Selected Second Order Stochastic Dominance Probabilities for Male and Female Subgroups**

| | Male | Female |
|---|---|---|
| $\Pr\left(2006_{SSD}2010\right)$ | 0.0562 | 0.1027 |
| $\Pr\left(2010_{SSD}2014\right)$ | 0.4740 | 0.4441 |
| $\Pr\left(2014_{SSD}2017\right)$ | 0.3792 | 0.4603 |
| $\Pr\left(2006_{SSD}2014\right)$ | 0.2457 | 0.3829 |
| $\Pr\left(2010_{SSD}2017\right)$ | 0.5905 | 0.8288 |
| $\Pr\left(2006_{SSD}2017\right)$ | 0.2822 | 0.7741 |

**Table 10: Dominance Probabilities for Male VS Female Subgroups for the Population with a Mental Health Score below 0.5**

| | 2001 | 2006 | 2010 | 2014 | 2017 |
|---|---|---|---|---|---|
| Pr(Female$_{FSD}$Male) | 0.0068 | 0.0004 | 0.0001 | 0.0004 | 0.0000 |
| Pr(Male$_{FSD}$Female) | 0.5059 | 0.3996 | 0.6483 | 0.8126 | 0.8962 |
| Pr(No FSD) | 0.4873 | 0.6000 | 0.3516 | 0.1870 | 0.1038 |
| Pr(Female$_{SSD}$Male) | 0.0419 | 0.0107 | 0.0010 | 0.0017 | 0.0008 |
| Pr(Male$_{SSD}$Female) | 0.5410 | 0.4177 | 0.6625 | 0.8275 | 0.9142 |
| Pr(No SSD) | 0.4171 | 0.5716 | 0.3365 | 0.1708 | 0.0850 |

**Table 11: Selected Dominance Probabilities for Male and Female Subgroups with a Mental Health Score below 0.5**

| | FSD | | SSD | |
|---|---|---|---|---|
| | Male | Female | Male | Female |
| Improving mental health | | | | |
| $\Pr\left(2006_{DOM}\,2001\right)$ | 0.4293 | 0.2819 | 0.4909 | 0.4973 |
| $\Pr\left(2010_{DOM}\,2006\right)$ | 0.4393 | 0.1771 | 0.5915 | 0.3208 |
| $\Pr\left(2010_{DOM}\,2001\right)$ | 0.7539 | 0.3738 | 0.7763 | 0.5821 |
| Deteriorating mental health | | | | |
| $\Pr\left(2010_{DOM}\,2014\right)$ | 0.5259 | 0.5409 | 0.5537 | 0.7234 |
| $\Pr\left(2014_{DOM}\,2017\right)$ | 0.3262 | 0.4215 | 0.3963 | 0.4759 |
| $\Pr\left(2010_{DOM}\,2017\right)$ | 0.5854 | 0.8254 | 0.5934 | 0.8461 |

**Table 12: Summary Statistics for the Health Scores for Indigenous and Non-indigenous Subgroups**

| | 2001 | | 2006 | | 2010 | | 2014 | | 2017 | |
|---|---|---|---|---|---|---|---|---|---|---|
| | indig | Non-ind | indig | Non-ind | indig | Non-ind | indig | Non-ind | indig | Non-ind |
| Sample Mean | 0.7082 | 0.7393 | 0.6986 | 0.7477 | 0.7046 | 0.7468 | 0.6654 | 0.7392 | 0.6716 | 0.7293 |
| Minimum | 0.1200 | 0.0400 | 0.0800 | 0.0400 | 0.1200 | 0.0400 | 0.0800 | 0.0400 | 0.0400 | 0.0400 |
| Maximum | 0.9990 | 0.9990 | 0.9990 | 0.9990 | 0.9990 | 0.9990 | 0.9990 | 0.9990 | 0.9990 | 0.9990 |
| Standard Deviation | 0.1802 | 0.1728 | 0.1952 | 0.1692 | 0.1774 | 0.1665 | 0.2182 | 0.1747 | 0.1929 | 0.1797 |
| Sample Size | 207 | 9515 | 206 | 8943 | 247 | 9298 | 379 | 11725 | 383 | 12190 |
| Headcount (cutoff = 0.5) | 0.1350 | 0.0992 | 0.1521 | 0.0904 | 0.1249 | 0.0928 | 0.2191 | 0.1032 | 0.1826 | 0.1172 |
| $FGT_1$ (cutoff = 0.5) | 0.0315 | 0.0252 | 0.0413 | 0.0221 | 0.0289 | 0.0706 | 0.0706 | 0.0267 | 0.0472 | 0.0301 |
| $FGT_2$ (cutoff = 0.5) | 0.0122 | 0.0108 | 0.0180 | 0.0094 | 0.0123 | 0.0347 | 0.0347 | 0.0113 | 0.0205 | 0.0129 |
| Proportion below 0.1 | 0.0000 | 0.0023 | 0.0024 | 0.0018 | 0.0000 | 0.0009 | 0.0057 | 0.0018 | 0.0058 | 0.0023 |

**Table 13: Dominance Probabilities for Indigenous VS Non-indigenous Subgroups**

| | 2001 | | 2006 | | 2010 | | 2014 | | 2017 | |
|---|---|---|---|---|---|---|---|---|---|---|
| | $y<1$ | $y\leq 0.5$ | $y<1$ | $y\leq 0.5$ | $y<1$ | $y\leq 0.5$ | $y<1$ | $y\leq 0.5$ | $y<1$ | $y\leq 0.5$ |
| Pr(ind$_{FSD}$non-ind) | 0.0029 | 0.0756 | 0.0004 | 0.0103 | 0.0001 | 0.0336 | 0.0000 | 0.0000 | 0.0000 | 0.0032 |
| Pr(Non-ind$_{FSD}$ind) | 0.1094 | 0.3978 | 0.0684 | 0.6965 | 0.1112 | 0.5727 | 0.3845 | 0.9470 | 0.1788 | 0.6172 |
| Pr(No FSD) | 0.8877 | 0.5266 | 0.9312 | 0.2932 | 0.8887 | 0.3937 | 0.6155 | 0.0530 | 0.8212 | 0.3796 |
| Pr(ind$_{SSD}$non-ind) | 0.0235 | 0.1725 | 0.0026 | 0.0314 | 0.0017 | 0.0801 | 0.0000 | 0.0001 | 0.0001 | 0.0306 |
| Pr(non-ind$_{SSD}$ind) | 0.4240 | 0.4358 | 0.7121 | 0.7212 | 0.6251 | 0.6315 | 0.9479 | 0.9480 | 0.6469 | 0.6470 |
| Pr(No SSD) | 0.5525 | 0.3917 | 0.2853 | 0.2474 | 0.3732 | 0.2884 | 0.0521 | 0.0519 | 0.3530 | 0.3224 |

**Table 14: Selected Dominance Probabilities for the Indigenous Subgroup**

| | Whole population | | Population with $y \leq 0.5$ | |
|---|---|---|---|---|
| | FSD | SSD | FSD | SSD |
| $\Pr(2006_{DOM}\,2001)$ | 0.0452 | 0.1254 | 0.1314 | 0.1901 |
| $\Pr(2001_{DOM}\,2006)$ | 0.0705 | 0.3788 | 0.4239 | 0.5087 |
| $\Pr(2006_{NO-DOM}\,2001)$ | 0.8843 | 0.4958 | 0.4447 | 0.3012 |
| $\Pr(2010_{DOM}\,2006)$ | 0.0549 | 0.3377 | 0.4115 | 0.4769 |
| $\Pr(2006_{DOM}\,2010)$ | 0.0454 | 0.1204 | 0.1144 | 0.1876 |
| $\Pr(2010_{NO-DOM}\,2006)$ | 0.8997 | 0.5419 | 0.4741 | 0.3355 |
| $\Pr(2014_{DOM}\,2010)$ | 0.0012 | 0.0097 | 0.0064 | 0.0157 |
| $\Pr(2010_{DOM}\,2014)$ | 0.1984 | 0.7298 | 0.7673 | 0.7814 |
| $\Pr(2014_{NO-DOM}\,2010)$ | 0.8004 | 0.2605 | 0.2263 | 0.2029 |
| $\Pr(2017_{DOM}\,2014)$ | 0.0708 | 0.4381 | 0.5774 | 0.6554 |
| $\Pr(2014_{DOM}\,2017)$ | 0.0092 | 0.0524 | 0.0360 | 0.0632 |
| $\Pr(2017_{NO-DOM}\,2014)$ | 0.9200 | 0.5096 | 0.3866 | 0.2814 |
| $\Pr(2017_{DOM}\,2001)$ | 0.0063 | 0.0400 | 0.0612 | 0.1305 |
| $\Pr(2001_{DOM}\,2017)$ | 0.2269 | 0.5248 | 0.5007 | 0.5514 |
| $\Pr(2017_{NO-DOM}\,2001)$ | 0.7668 | 0.4352 | 0.4381 | 0.3181 |

Figure 1: Estimated Health Density Functions

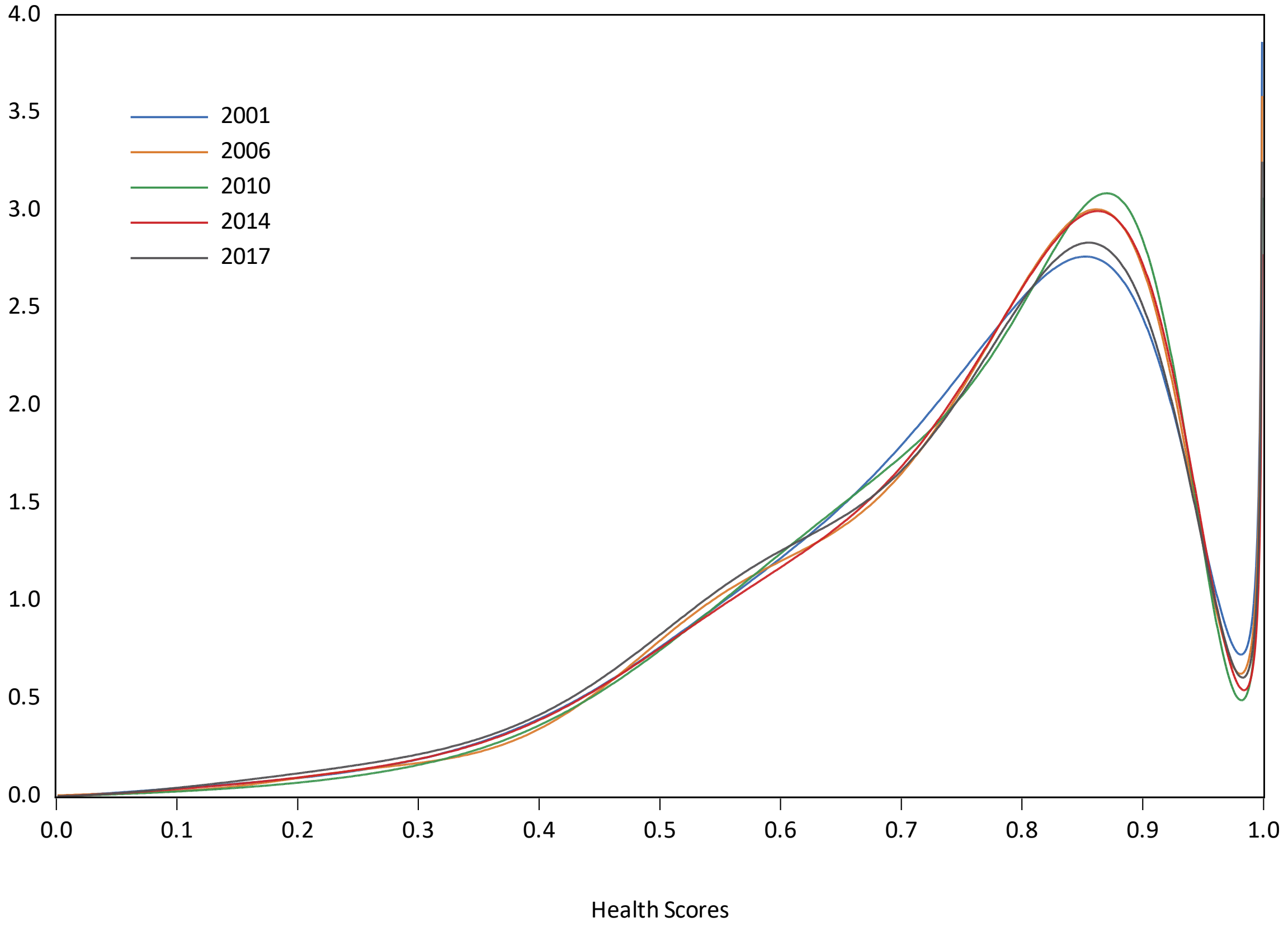

Figure 2: Posterior Density Functions for Mean Health

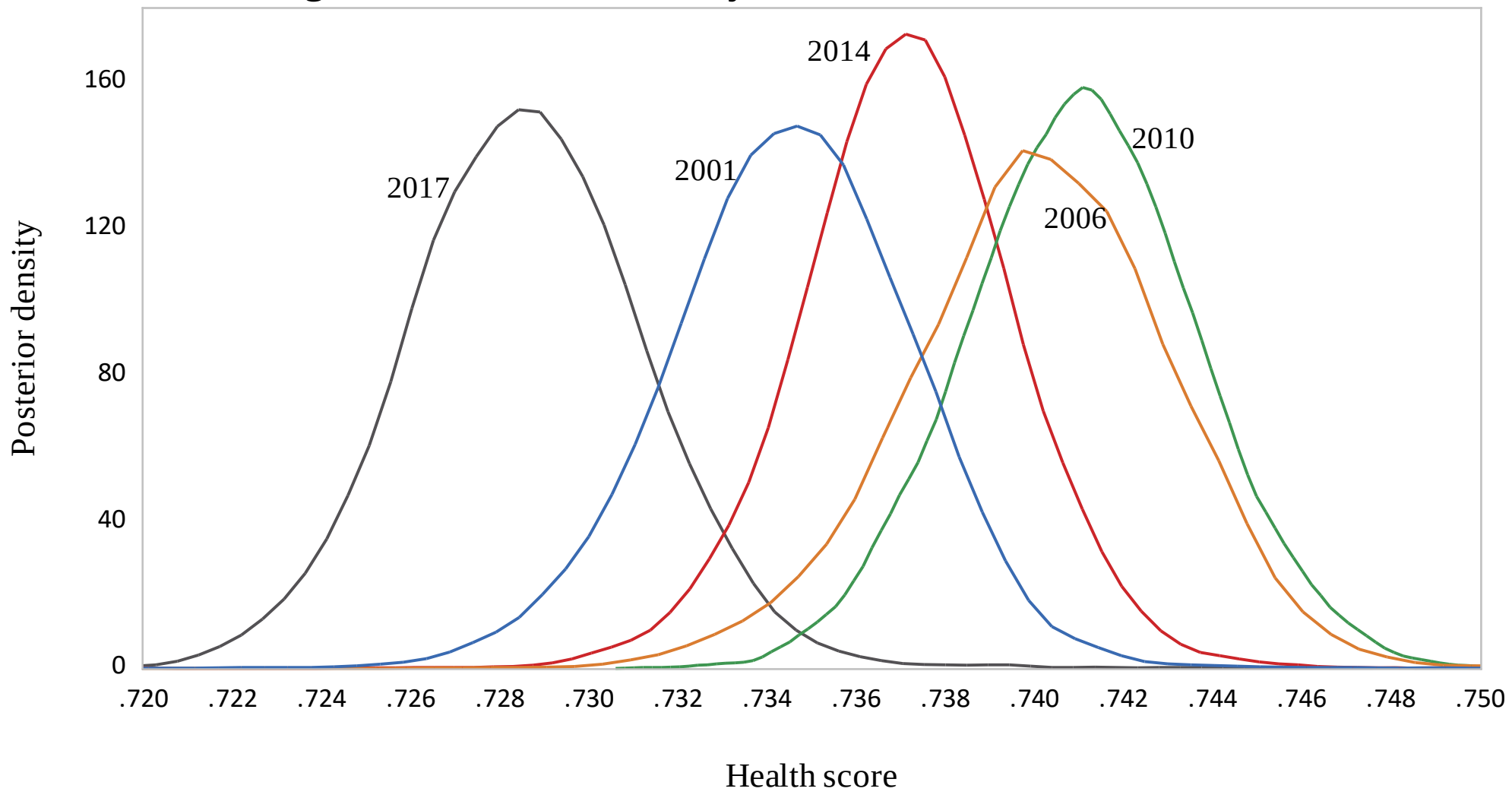

Figure 3: Estimated Distribution Functions.

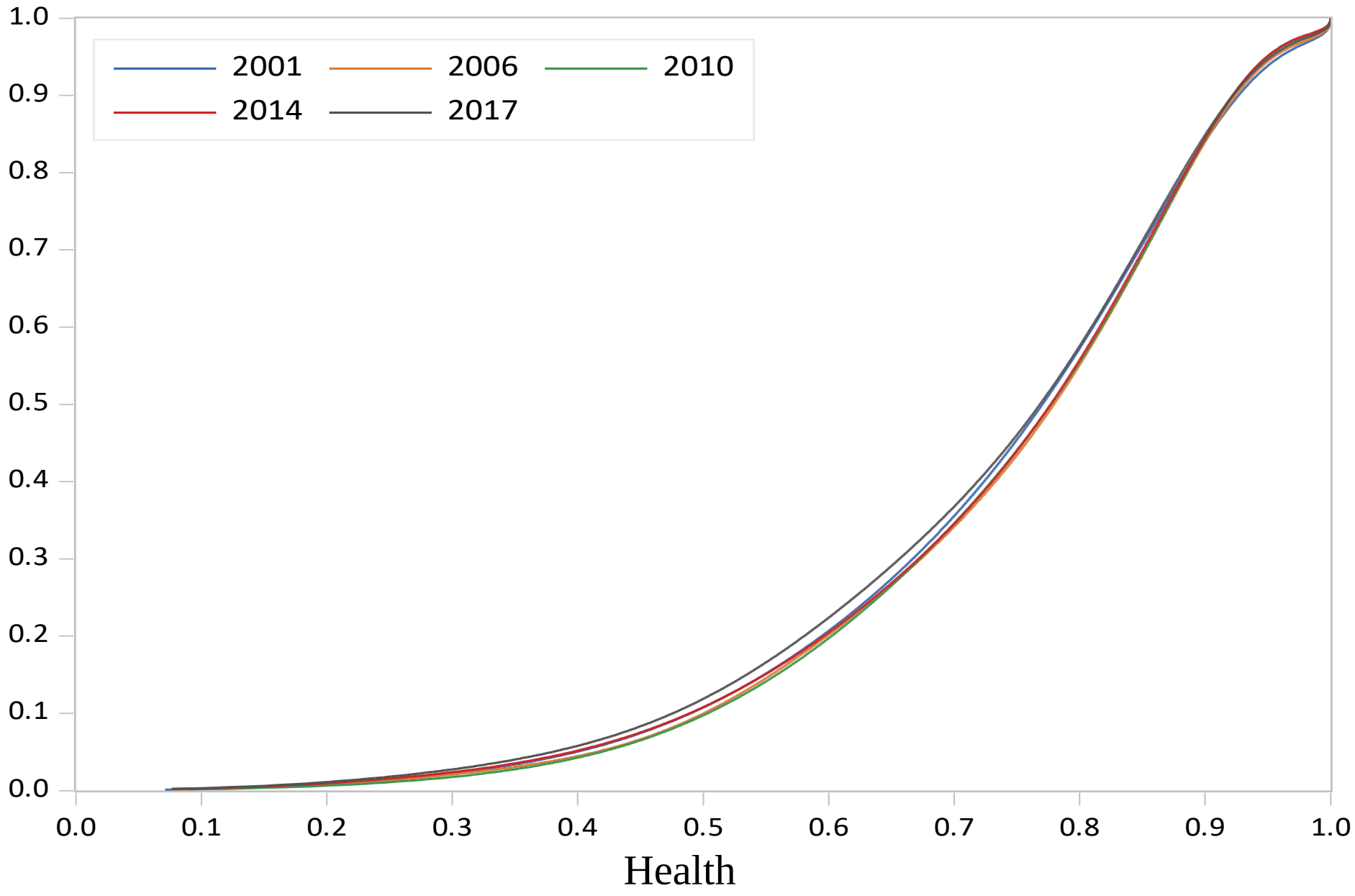

Figure 4: Dominance Probabilities and estimated Distribution Function Differences for (a) $2006_{FSD}2017$ and (b) $2010_{FSD}2017$.

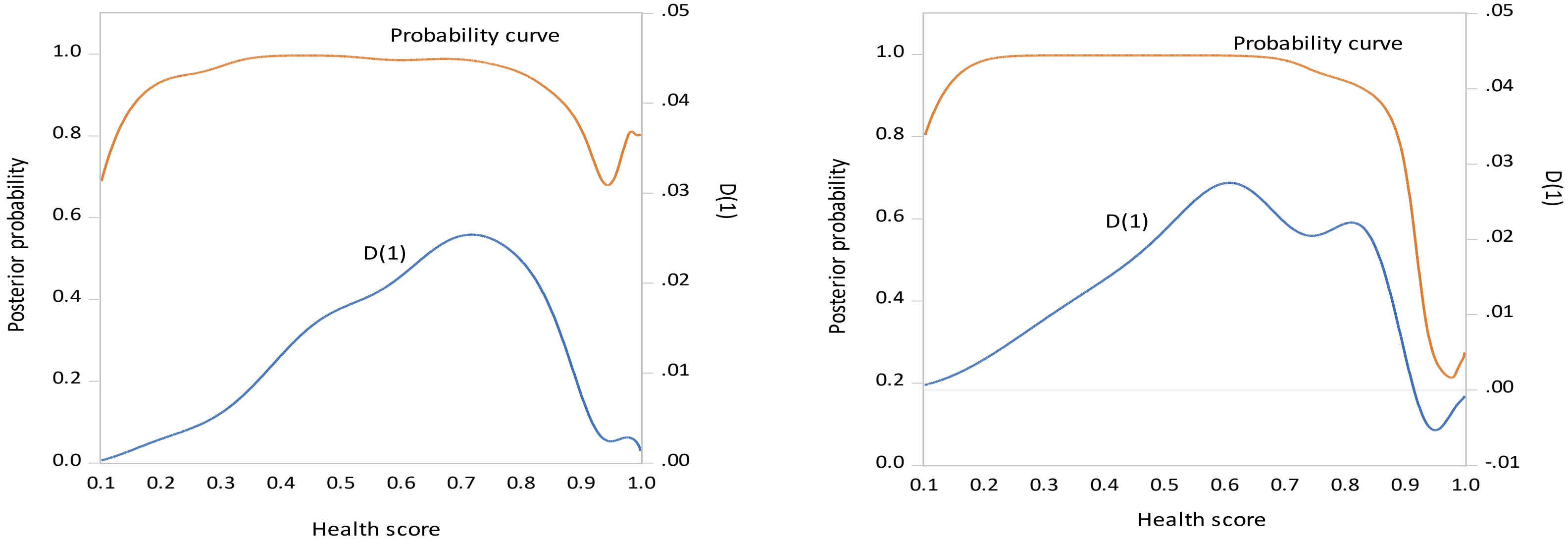

(a) (b)

Figure 5: Probability Curves and Function Differences
(a) $2006_{SSD}2017$ and (b) $2010_{SSD}2017$.

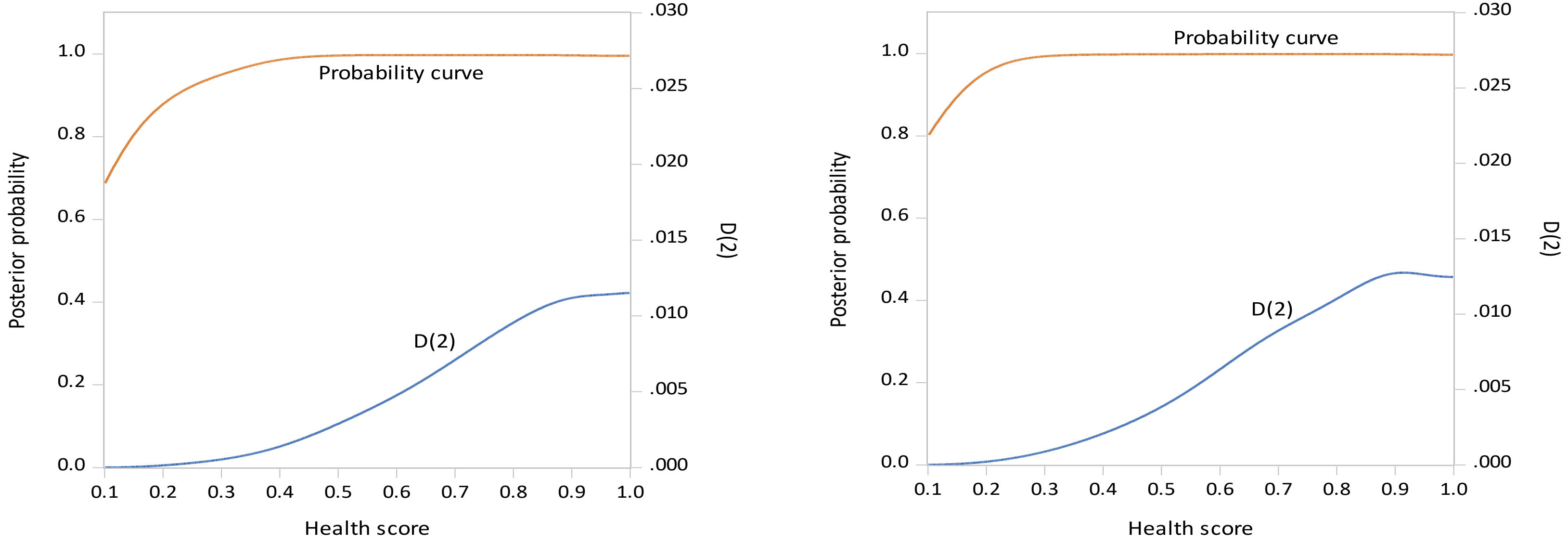

(a) (b)

Figure 6: Probability Curves and Function Differences
(a) $Male_{FSD}Female$ and (b) $Male_{SSD}Female$ in 2017

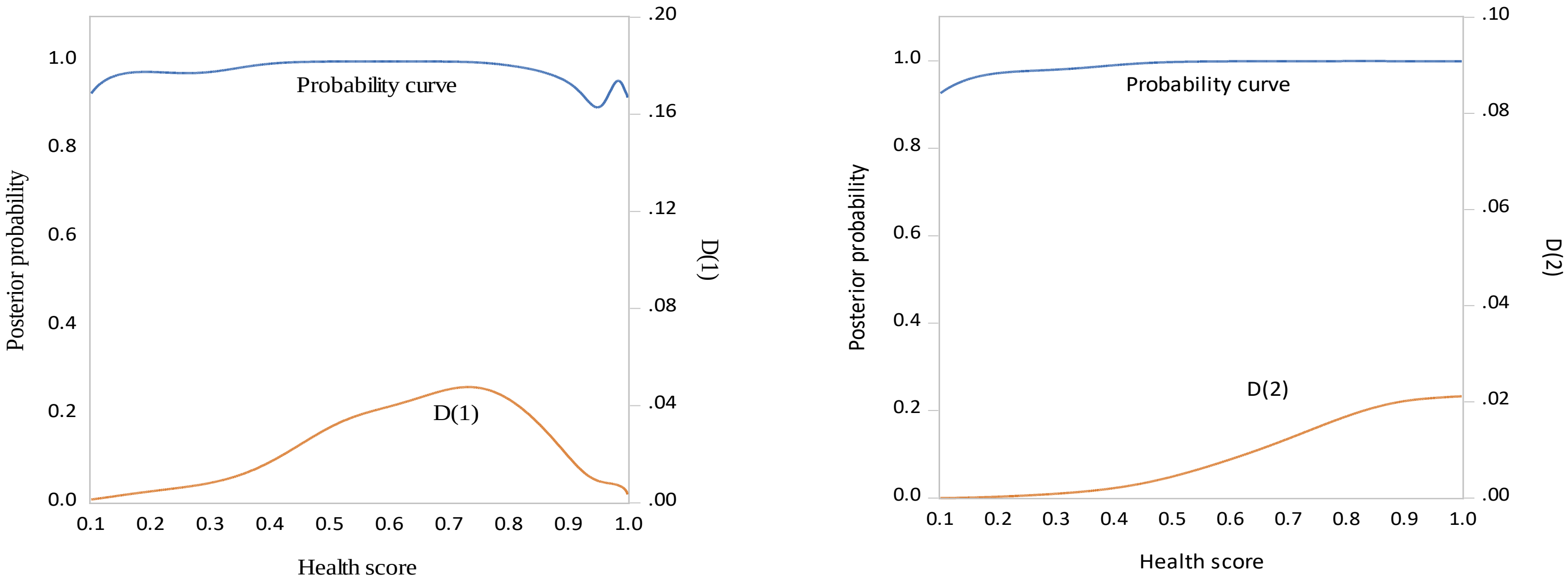

(a) (b)

Figure 7 Posterior densities for mean health score

for Indigenous and Non-indigenous populations in 2001 and 2017.

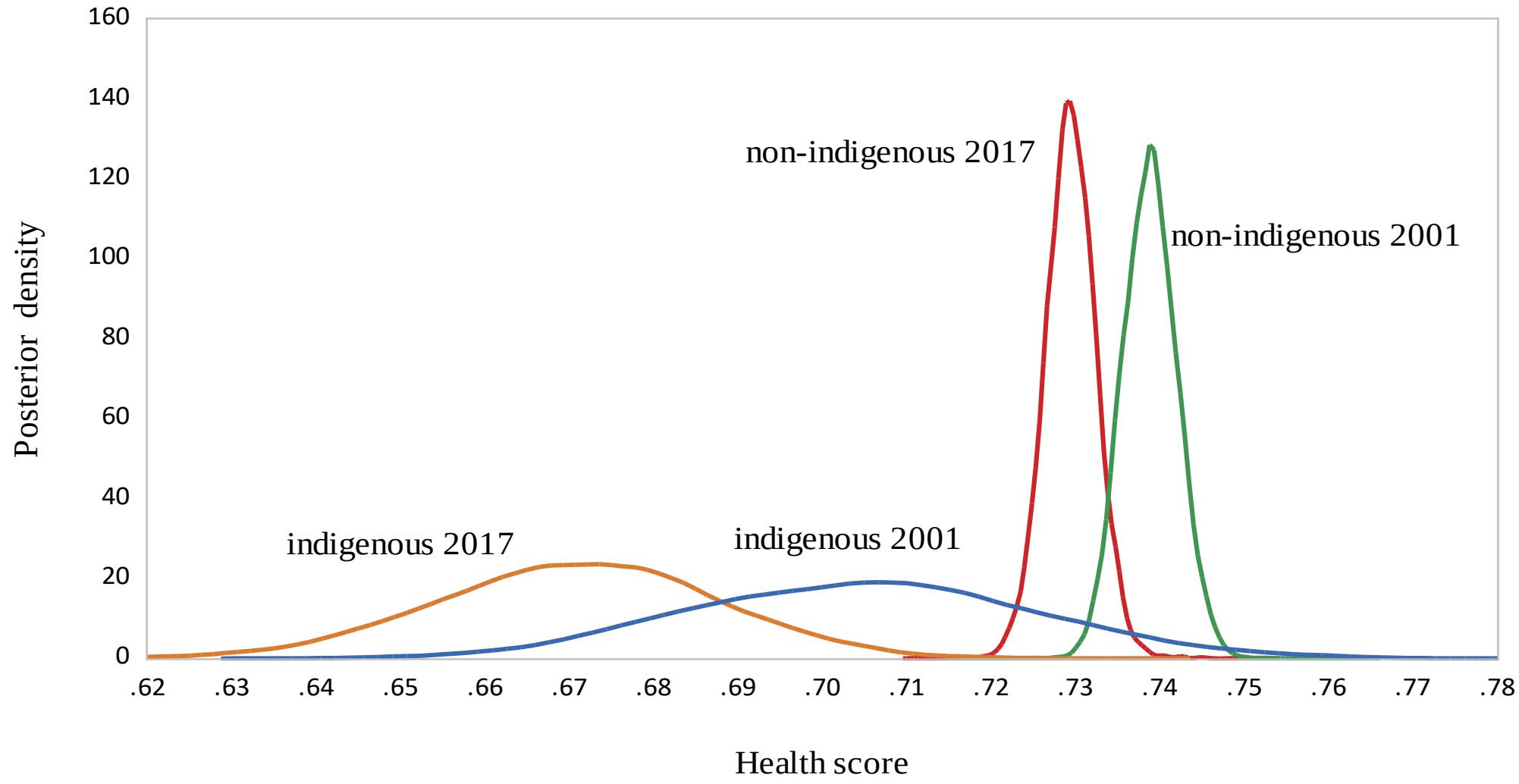

Figure 8: Posterior densities for proportion of people with poor mental health for

Indigenous and Non-indigenous populations in 2001 and 2017

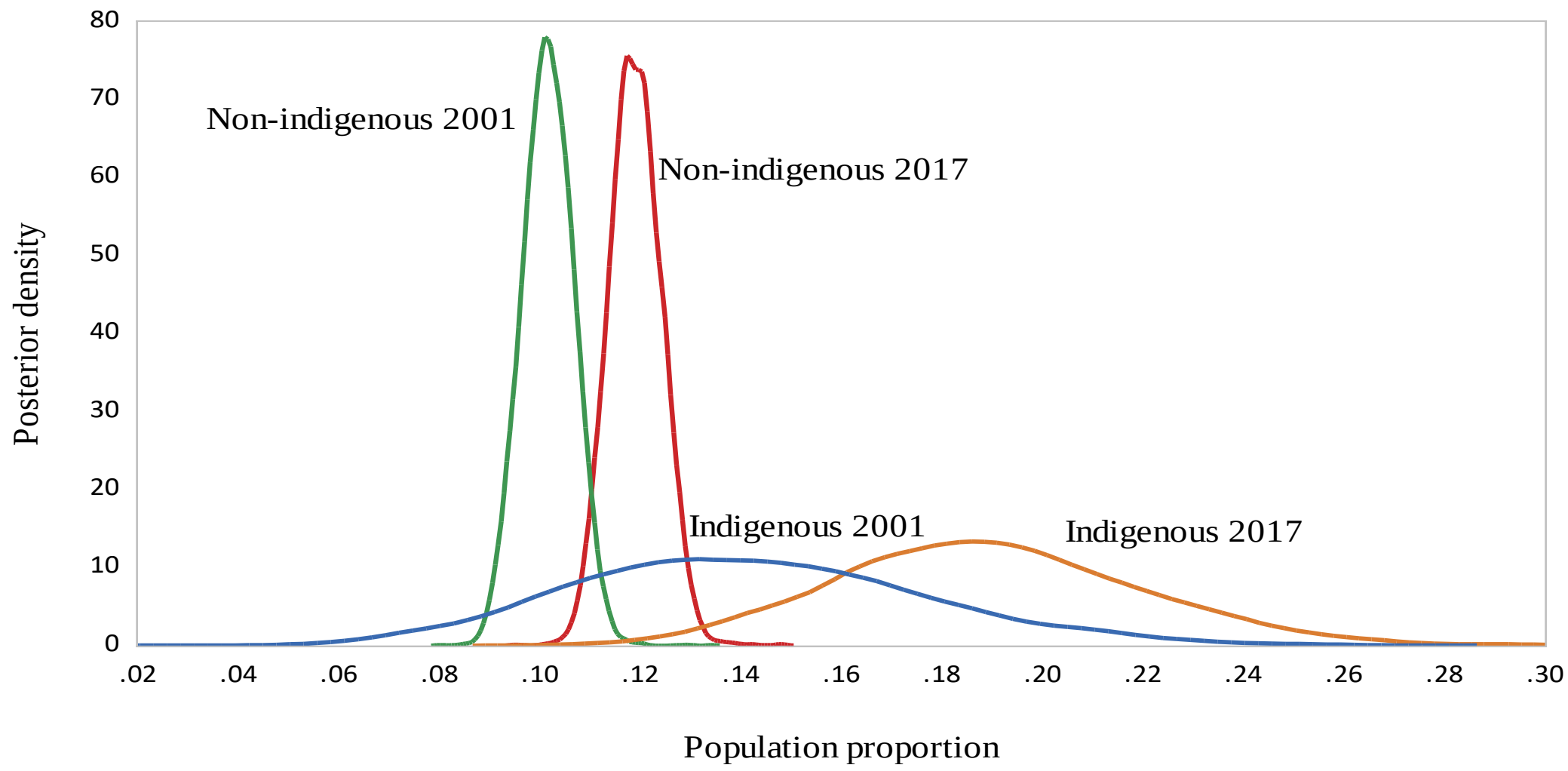

Figure 9: Probability Curves and Function Differences for the Non-indigenous Dominating the Indigenous Distribution

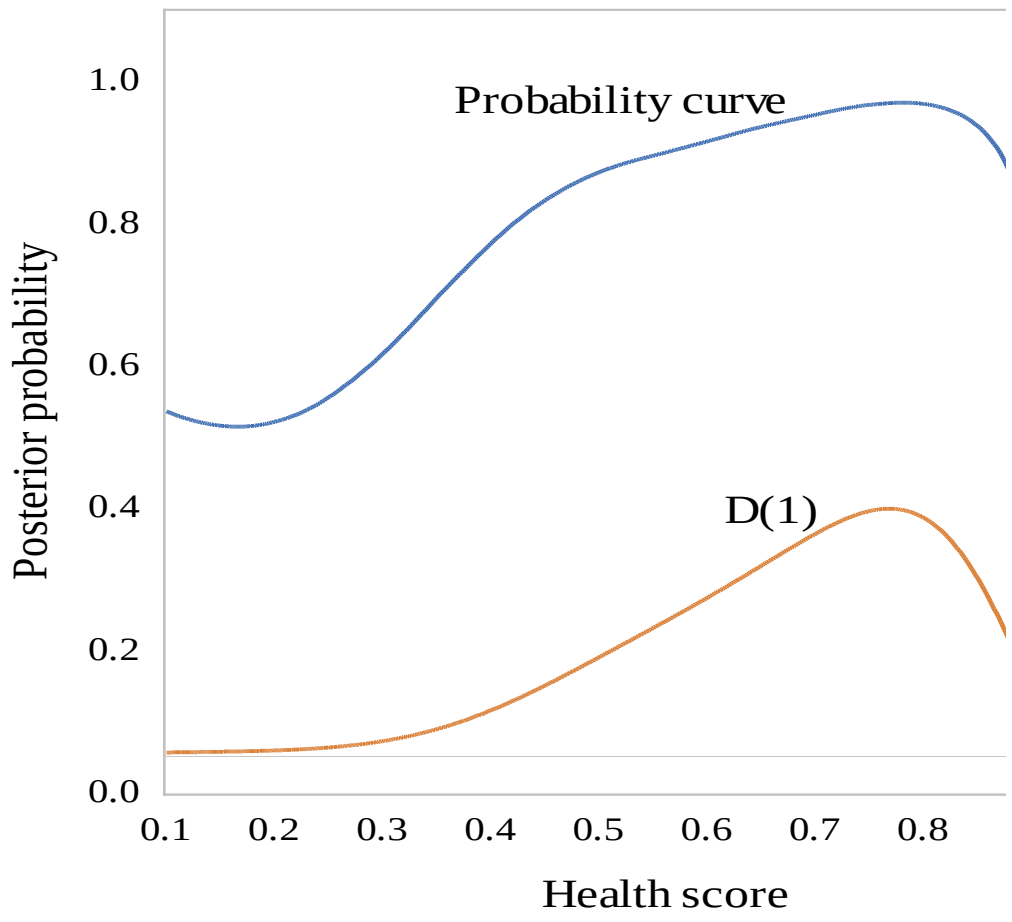

(a) FSD in 2001

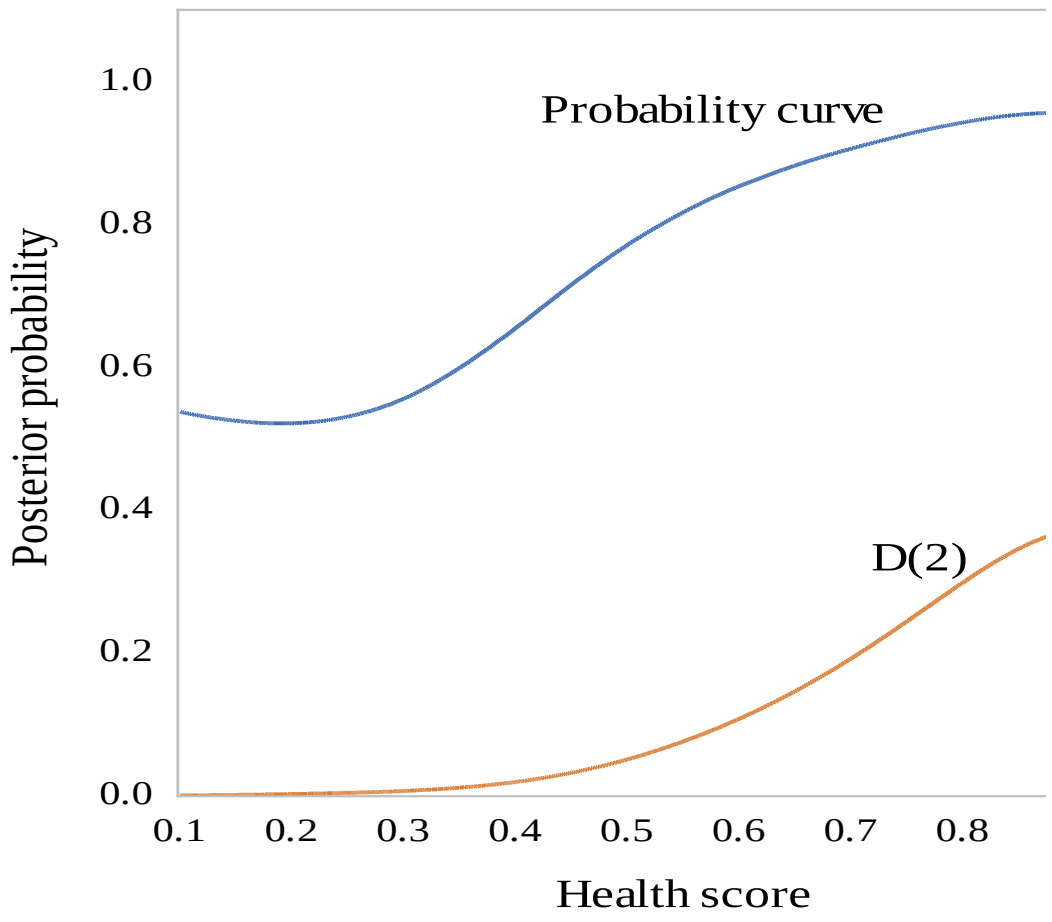

(b) SSD in 2001

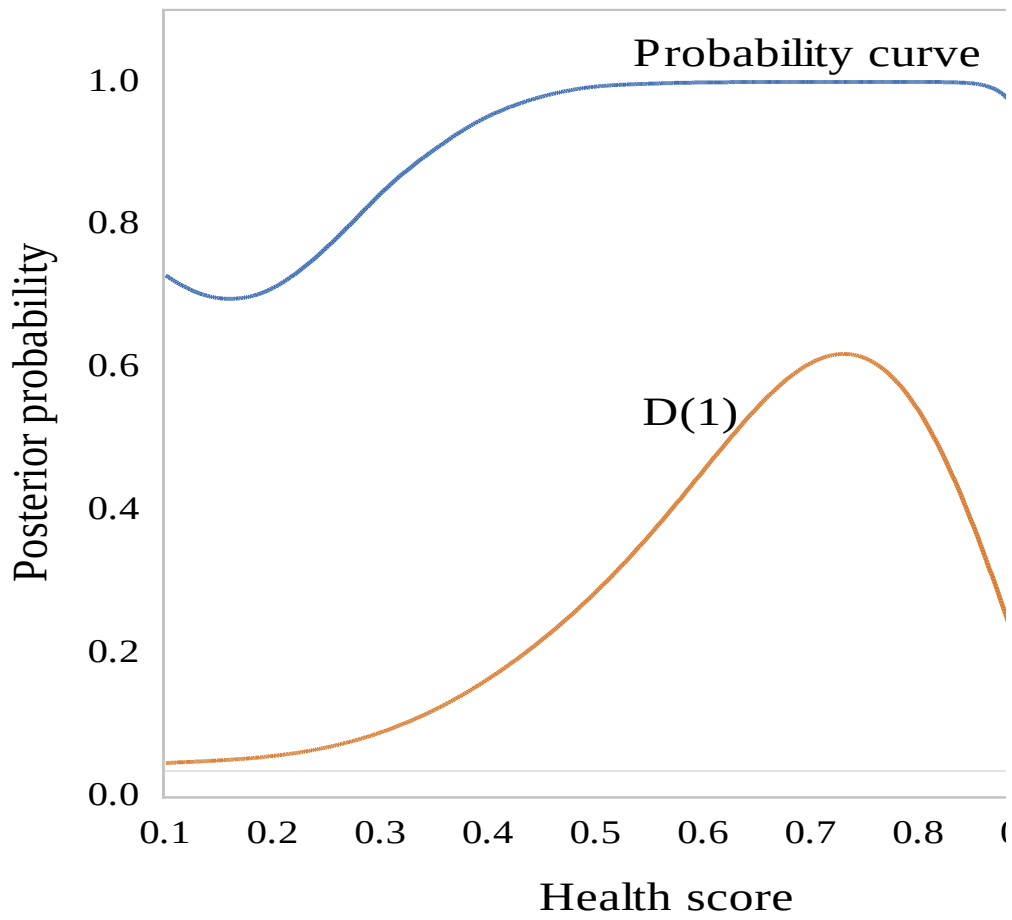

(c) FSD in 2017

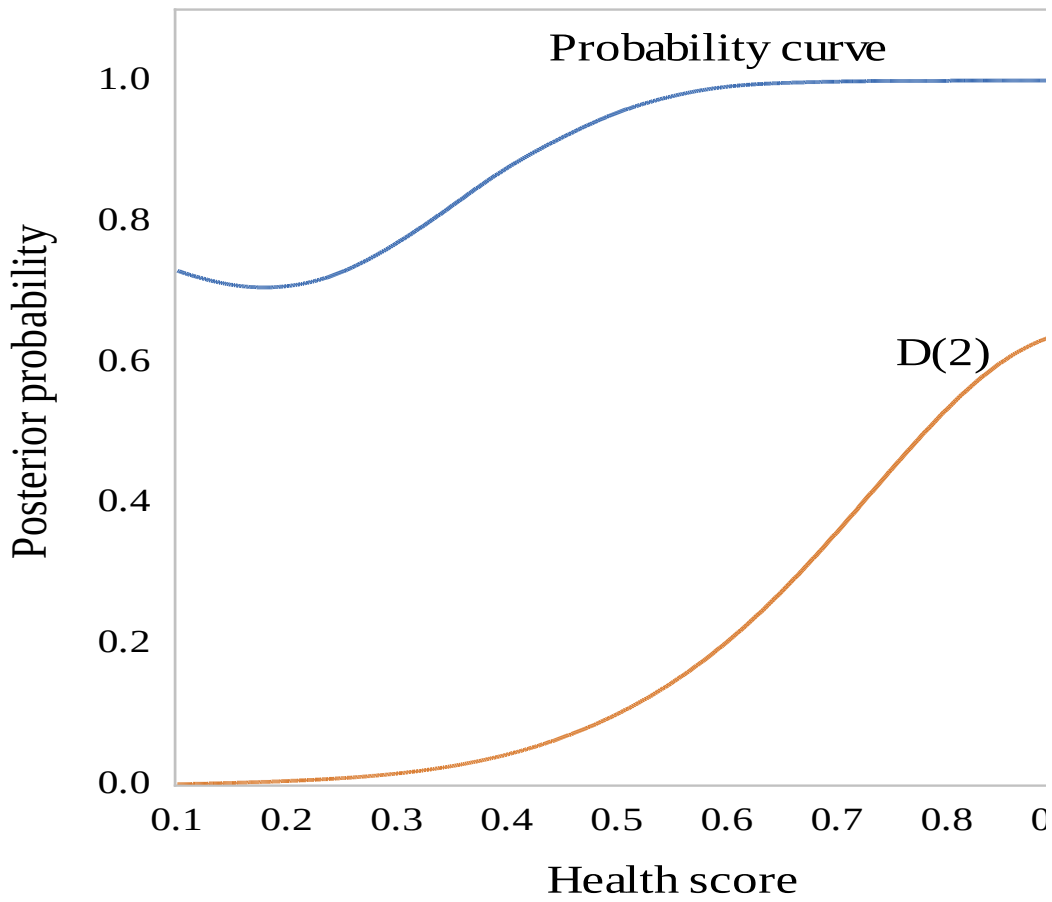

(d) SSD in 2017

# Comparisons of Australian Mental Health Distributions

David Gunawan
*University of Wollongong*

William Griffiths*
*University of Melbourne*

Duangkamon Chotikapanich
*Monash University*

10 June, 2021

## Supplementary Appendix

**A. Estimating the Beta Mixture Model**

To estimate the parameters in equation (1), it is convenient to reparameterize it as

$$p(y\,|\,\boldsymbol{m},\boldsymbol{s},\boldsymbol{w}) = \sum_{k=1}^{\infty} w_k B(y\,|\,m_k, s_k)$$

where $m_k = \alpha_k / (\alpha_k + \beta_k)$ is the mean of the *k*-th component, $s_k = \alpha_k + \beta_k$, $\boldsymbol{m}$ and $\boldsymbol{s}$ are vectors containing these elements, and

$$B(y\,|\,s_k, m_k) = \frac{\Gamma(s_k)}{\Gamma(s_k m_k)\Gamma(s_k(1-m_k))} y^{s_k m_k - 1}(1-y)^{s_k(1-m_k)-1}$$

Prior distributions known as base distributions are placed on the parameters $m_k$ and $s_k$; we denote them by $G_{0,m}$ and $G_{0,s}$, respectively. We assume that $G_{0,m}$ is a beta distribution, $B\left(m_k, \tilde{\alpha}, \tilde{\beta}\right)$, and $G_{0,s}$ is an exponential distribution with parameter $\lambda$, $p\left(s_k\,|\,\lambda\right) = \lambda \exp\left(-\lambda s_k\right)$. For a prior distribution on the weights $w_k$, we use what is known as a stick-breaking process, $w_k \sim SBP(\alpha_0)$, where

$$w_k = \eta_k \prod_{j<k} (1-\eta_j)$$

with the $\eta_k$ being draws from the beta distribution $\eta_k \sim B(1, \alpha_0)$. The parameters $\left(\tilde{\alpha}, \tilde{\beta}, \lambda\right)$ are prior parameters whose values are set by the investigator. A hierarchical gamma prior $\alpha_0 \sim G(\underline{\alpha}, \underline{\beta})$ is used for $\alpha_0$ with $(\underline{\alpha}, \underline{\beta})$ being set by the investigator. The intuition behind the stick-breaking process is as follows. We have a stick of unit length from which probabilities $w_k$ are to be allocated to each component of the mixture. The first probability $w_i = \eta_1$ is drawn from a beta distribution with parameters $\left(1, \alpha_0\right)$. There is then $1-\eta_1$ of the stick remaining to be allocated. To obtain a second value $w_2$, we break off a proportion $\eta_2 \sim B\left(1, \alpha_0\right)$ from the stick of length $1-\eta_1$ and set $w_2 = \eta_2\left(1-\eta_1\right)$. As we proceed, the stick gets shorter and shorter, and the $w_k$ decrease stochastically. The effect of these assumptions is to create a posterior distribution for the weights which is a Dirichlet distribution with parameters that are a weighted average of the proportion of observations allocated to each component of the mixture and the prior parameters implied by the base distribution $G_0 = \left(G_{0,m}, G_{0,s}\right)$. The prior parameter $\alpha_0$ is called a concentration parameter and controls the relative weight placed on $G_0$. Further details of the stick-breaking process can be found in Gelman et al. (2014, Ch.23).

Two latent variables are introduced to aid development of an MCMC algorithms. One is a uniform random variable $\tilde{u}_i$ used in the slice sampler proposed by Walker (2007). At each iteration, this device avoids the apparent necessity to sample an infinite number of the parameters $(m_k, s_k)$ by truncating the infinite number of components to a finite number $K$ that depends on draws $\tilde{u}_1, \tilde{u}_2, \ldots, \tilde{u}_n$, $n$ being the sample size. Posterior sampling proceeds using the finite number of components. A residual weight $w_{K+1} = 1 - \sum_{k=1}^{K} w_k$ ensures the mixture is complete. The value of $K$ can change with each iteration. A second latent variable $v_i$ identifies the component to which the *i*-th observation is assigned. Conditional on $v_i = k,$ the distribution of $y_i$ follows the beta distribution of the *k*-th component,

$$(y_i \mid v_i = k, \boldsymbol{m}, \boldsymbol{s}) \sim B(y_i \mid m_k, s_k)$$

and $\Pr(v_i = k \mid \boldsymbol{w}) = w_k.$ The specification can be summarized as:

$$w_k \sim SBP(\alpha_0)$$

$$m_k \sim G_{0,m} \quad \text{for } k = 1, 2, \ldots, \infty$$

$$s_k \sim G_{0,s} \quad \text{for } k = 1, 2, \ldots, \infty$$

$$(y_i \mid v_i = k) \sim B(y_i \mid m_k, s_k)$$

$$\Pr(v_i = k) = w_k$$

$$(\tilde{u}_i \mid v_i = k) \sim U(0, w_k)$$

Corresponding to the observed health outcomes $\boldsymbol{y} = (y_1, y_2, \ldots, y_n)$ are sampling weights $\boldsymbol{\tau} = (\tau_1, \tau_2, \ldots, \tau_n).$ The MCMC algorithm used to draw observations from the joint posterior density for $(\alpha_0, \tilde{\boldsymbol{u}}, \boldsymbol{v}, \boldsymbol{w}, \boldsymbol{m}, \boldsymbol{s})$ is applied to a series of pseudo-representative samples drawn using the Bayesian bootstrap procedure described in Gunawan et al. (2020). The framework of the MCMC steps is as follows.

1. Generate a pseudo-representative sample $\tilde{\boldsymbol{y}} = (\tilde{y}_1, \tilde{y}_2, \ldots, \tilde{y}_n)$.

2. Sample $\left(\tilde{\boldsymbol{u}} \mid \tilde{\boldsymbol{y}}, \boldsymbol{s}, \boldsymbol{w}, \boldsymbol{m}, \boldsymbol{v}, \alpha_0\right)$

3. Sample $\left(\boldsymbol{w} \mid \tilde{\boldsymbol{y}}, \tilde{\boldsymbol{u}}, \boldsymbol{s}, \boldsymbol{m}, \boldsymbol{v}, \alpha_0\right)$ and update $K$

4. Sample $\left(\boldsymbol{m}, \boldsymbol{s} \mid \tilde{\boldsymbol{y}}, \boldsymbol{w}, \boldsymbol{v}, \tilde{\boldsymbol{u}}, \alpha_0\right)$

5. Sample $\left(\boldsymbol{v} \mid \tilde{\boldsymbol{y}}, \boldsymbol{w}, \boldsymbol{m}, \boldsymbol{s}, \tilde{\boldsymbol{u}}, \alpha_0\right)$

6. Sample $\left(\alpha_0 \mid \boldsymbol{m}, \boldsymbol{s}, \boldsymbol{v}\right)$

7. Compute quantities of interest.

8. Repeat steps 2 to 7, $M$ times.

9. Repeat steps 1 to 8, $J$ times.

Details of each step follow.

Step 1

Let $N$ be the size of the population. We set $N = 600{,}000.$ The results are not sensitive to this setting as long as $N$ is much bigger than $n$. The sample $\left(y_1, y_2, \ldots, y_n\right)$ is augmented with $N-n$ values $\left(y_1^*, y_2^*, \ldots, y_{N-n}^*\right)$ to form a pseudo population. A random sample of size $n$, $\left(\tilde{y}_1, \tilde{y}_2, \ldots \tilde{y}_n\right)$ is then drawn from this pseudo population. Let $N^* = (N-n)/n.$ The sample weights are normalized such that $\sum_{i=1}^{n} \tau_i = N.$ Let $\ell_i = 0, i = 1, 2, \ldots, n.$ For $j = 1, 2, \ldots, N-n,$ draw $y_j^* = y_i$ with probability

$$\frac{\tau_i - 1 + \ell_i N^*}{N - n + (j-1)N^*}$$

Each time a $y_i$ is used the corresponding $\ell_i$ is incremented by 1, such that $\sum_{i=1}^{n} \ell_i = j-1.$

Initialisation

Before proceeding with the remaining steps, we need initial values for the parameters. The following initialisations are suitable.

(i) Set $K = 3$ as an initial number of finite components and allocate the data randomly to these three components. Associated with each observation $y_i$, $i = 1, 2, \ldots, n$, is a latent variable $v_i$, equal to 1,2 or 3, identifying the component to which $y_i$ has been allocated.

(ii) In each iteration, there is scope for increasing the number of categories. To distinguish between the total number of categories, and those which contain observations in the current iteration, we denote the number of categories that contain observations by $M^*$. Set $M^* = K = 3$.

(iii) Generate $\left(w_1, w_2, w_3\right)$ from a Dirichlet (1,1,1) distribution.

(iv) Sample $m_k$ and $s_k$ from their prior distributions for $k = 1, 2, 3$.

Step 2

Sample $\tilde{u}_i$ from a uniform density on the interval $\left(0, w_{v_i}\right)$ for $i = 1, 2, \ldots, n.$

Step 3(a)

Sample $\boldsymbol{w} = \left(w_1, w_2, \ldots, w_K, w_{K+1}\right)$ from a Dirichlet distribution $D\left(n_1, n_2, \ldots, n_K, \alpha_0\right)$ with $n_k$ being the number of observations for which $v_i = k.$ The weight $w_{K+1}$ is a residual weight whose corresponding parameter is $\alpha_0$. In the initial iteration $\alpha_0$ is set equal to its prior mean.

Step 3(b)

(i) If $w_{K+1} \le \min\left(\tilde{u}_1, \tilde{u}_2, \ldots, \tilde{u}_n\right)$, go to Step 4.

(ii) If $w_{K+1} > \min\left(\tilde{u}_1, \tilde{u}_2, \ldots, \tilde{u}_n\right)$, break the residual weight $w_{K+1}$ into two pieces; draw $\eta \sim B(1, \alpha_0)$, set $w_{K+2} = (1-\eta) w_{K+1}$ and change $w_{K+1}$ to $\eta w_{K+1}$. Sample new values $m_{K+1}$ from its base distribution $B(\tilde{\alpha}, \tilde{\beta})$, and $s_{K+1}$ from its base distribution, exponential with parameter $\lambda.$ Set $K = K + 1$, and go back to Step 3(b)(i).

Step 4

Values for $m_k$ and $s_k$ that may have been sampled from their priors in Step 3(b)(ii) were for potentially new categories which do not contain any observations in the current iteration. In Step 4 we sample $m_k$ and $s_k$ for all categories that contain observations in the current iteration. That is, draw values $m_k$ and $s_k$ for $k = 1, 2, \ldots, M^*,$ from

$$p\left(m_k, s_k \mid \tilde{\boldsymbol{y}}, \boldsymbol{v}, \boldsymbol{w}\right) \propto p(s_k)\, p(m_k) \prod_{v_i = k} B\left(\tilde{y}_i \mid m_k, s_k\right)$$

This density is not a recognizable form and requires a Metropolis-Hastings (M-H) step. The M-H steps can be described as follows:

1. Generate $\left(s_k^*, m_k^*\right)$ from a proposal density $q\left(s_k, m_k \mid s_k^{(t-1)}, m_k^{(t-1)}\right)$ and a constant $c$ from a uniform (0,1) distribution.
2. Compute the acceptance probability

$$r = \frac{p\left(m_k^*, s_k^* \mid \tilde{\boldsymbol{y}}, \boldsymbol{v}, \boldsymbol{w}\right) q\left(s_k^{(t-1)}, m_k^{(t-1)} \mid s_k^*, m_k^*\right)}{p\left(m_k^{(t-1)}, s_k^{(t-1)} \mid \tilde{\boldsymbol{y}}, \boldsymbol{v}, \boldsymbol{w}\right) q\left(s_k^*, m_k^* \mid s_k^{(t-1)}, m_k^{(t-1)}\right)}$$

3. If $c < r$, then $\left(s_k^{(t)}, m_k^{(t)}\right) = \left(s_k^*, m_k^*\right)$, otherwise $\left(s_k^{(t)}, m_k^{(t)}\right) = \left(s_k^{(t-1)}, m_k^{(t-1)}\right)$.

A major issue with this algorithm is the need to choose the proposal distribution $q$. Both $s_k$ and $m_k$ are constrained parameters, i.e., $s_k > 0$ and $0 < m_k < 1$. This can result in poor mixing if many candidate parameters are proposed which violate these constraints, increasing the autocorrelation in the Markov chain. One approach to improve mixing is to reparameterize the constrained parameters to obtain unconstrained parameters, which assume any value on the real line. A suitable reparameterization is $s_k = \exp\left(\xi_k\right)$ and $m_k = \text{logit}(\kappa_k)$ such that $\xi_k$ and $\kappa_k$ are unconstrained parameters. We use random walk proposals for both $\xi_k$ and $\kappa_k$,

$$\xi_k^* = \xi_k^{(t-1)} + \nu_\xi \varepsilon$$

$$\kappa_k^* = \kappa_k^{(t-1)} + \nu_\kappa \varepsilon$$

where $\varepsilon$ is a draw from a standard normal distribution. We follow Garthwaite et al. (2016) to adaptively tune the scaling factors $\nu_\xi$ and $\nu_\kappa$ to achieve a pre-specified acceptance probability. Since the number of components at each iteration is random, we set the overall target acceptance probability to 25% across all mixture components $k$. The acceptance probability needs to be modified to include the Jacobians of the transformations. It becomes

$$r = \frac{p\left(m_k^*, s_k^* \mid \tilde{\boldsymbol{y}}, \tilde{\boldsymbol{s}}, \boldsymbol{w}\right)}{p\left(m_k^{(t-1)}, s_k^{(t-1)} \mid \tilde{\boldsymbol{y}}, \tilde{\boldsymbol{s}}, \boldsymbol{w}\right)} \left|\frac{s_k^*}{s_k}\right| \left|\frac{\dfrac{1}{m} + \dfrac{1}{1-m}}{\dfrac{1}{m^*} + \dfrac{1}{1-m^*}}\right|$$

Step 5

In this step, we reallocate observations to components. For each $v_i$, we draw a value for $k$ from $k = \{1, 2, \ldots, K\}$ with probability

$$\Pr\left(v_i = k \mid K, \boldsymbol{m}, \boldsymbol{s}, \boldsymbol{w}, \tilde{\boldsymbol{u}}, \tilde{\boldsymbol{y}}\right) = \frac{I\left(\tilde{u}_i < w_k\right) B\left(y_i \mid m_k, s_k\right)}{\sum_{h=1}^{K} I\left(\tilde{u}_i < w_h\right) B\left(y_i \mid m_h, s_h\right)} \qquad \text{for } i = 1, 2, \ldots, n.$$

where $I\left(\tilde{u}_i < w_k\right)$ is an indicator function; $I\left(\tilde{u}_i < w_k\right) = 1$ when $\tilde{u}_i < w_k$, otherwise $I\left(\tilde{u}_i < w_k\right) = 0$. All components with no observations allocated to them are removed and both $K$ and $M^*$ are reset to the number of components with a positive number of observations. The values $\left(w_k, m_k, s_k\right)$, $k = 1, 2, \ldots K$, are rearranged accordingly.

Step 6

Draw $\alpha_0$ using the following steps.

(i) Draw $x \mid \alpha_0$ from $B\left(\alpha_0, n\right)$.

(ii) Draw $\alpha_0 \mid x$ from $G\left(\underline{\alpha} - \log(x), \underline{\beta} + K\right)$.

Step 7

Compute the residual weight $w_{K+1} = 1 - \sum_{k=1}^{K} w_k$. Sample new values $m_{K+1}$ from its base distribution $B(\tilde{\alpha}, \tilde{\beta})$, and $s_{K+1}$ from its base distribution, exponential with parameter $\lambda$. Then, after setting $\alpha_k = s_k m_k$ and $\beta_k = s_k (1 - m_k)$, compute draws on the following quantities of interest.

Mean, $$\mu = \sum_{k=1}^{K+1} w_k m_k$$

Second moment for the $k$-th component, $$\mu_{2,k} = \frac{\alpha_k\left(\alpha_k + 1\right)}{\left(\alpha_k + \beta_k\right)\left(\alpha_k + \beta_k + 1\right)}$$

Second moment, $$\mu_2 = \sum_{k=1}^{K+1} w_k \mu_{2,k}$$

Density function for any value $0 < y < 1$, $$p(y \mid \boldsymbol{\alpha}, \boldsymbol{\beta}, \boldsymbol{w}) = \sum_{k=1}^{K+1} w_k B\left(y \mid \alpha_k, \beta_k\right)$$

Cumulative distribution function for any value $0 < y < 1$, $$F(y \mid \boldsymbol{\alpha}, \boldsymbol{\beta}, \boldsymbol{w}) = \sum_{k=1}^{K+1} w_k F\left(y \mid \alpha_k, \beta_k\right)$$

First moment distribution function for any value $0 < y < 1$,

$$F^{(1)}(y \mid \boldsymbol{\alpha}, \boldsymbol{\beta}, \boldsymbol{w}) = \frac{1}{\mu} \sum_{k=1}^{K+1} w_k m_k F\left(y \mid \alpha_k + 1, \beta_k\right)$$

Second moment distribution function for any value $0 < y < 1$,

$$F^{(2)}(y \mid \boldsymbol{\alpha}, \boldsymbol{\beta}, \boldsymbol{w}) = \frac{1}{\mu_2} \sum_{k=1}^{K+1} w_k \mu_{2,k} F\left(y \mid \alpha_k + 2, \beta_k\right)$$

Headcount, $HC = F(z \mid \boldsymbol{\alpha}, \boldsymbol{\beta}, \boldsymbol{w})$

FGT(1), $FGT_1 = HC - \frac{\mu}{z} F^{(1)}(z \mid \boldsymbol{\alpha}, \boldsymbol{\beta}, \boldsymbol{w})$

FGT(2), $FGT_2 = HC - \frac{2\mu}{z} F^{(1)}(z \mid \boldsymbol{\alpha}, \boldsymbol{\beta}, \boldsymbol{w}) + \frac{\mu_2}{z^2} F^{(2)}(z \mid \boldsymbol{\alpha}, \boldsymbol{\beta}, \boldsymbol{w})$

**Prior Settings and Number of Draws**

The prior parameters we used in our empirical work were $\underline{\alpha} = \underline{\beta} = 10$, $\tilde{\alpha} = \tilde{\beta} = 2$ and $\lambda = 0.1$. These prior densities cover a large range of possible values and are non-informative. Probability intervals with 95% coverage are (0.0934, 0.9051) for $m_k$, (0.2285, 37.49) for $s_k$, and (0.4828, 1.7080) for $\alpha_0$. We generated 200 pseudo representative samples, and, for each of these samples, we obtained a total of 6000 MCMC draws. The first 1,000 of the MCMC draws were discarded as a burn in, and every 100th draw of the remaining 5,000 draws was retained. This strategy gave a total of 10,000 draws for dominance assessment.

## B. Extra tables

**Table S1: Posterior Means (Standard Deviations) for Mean Health Scores for Males and Females**

| | 2001 | | 2006 | | 2010 | | 2014 | | 2017 | |
|---|---|---|---|---|---|---|---|---|---|---|
| | Male | Female | Male | Female | Male | Female | Male | Female | Male | Female |
| Mean | 0.7467 | 0.7240 | 0.7526 | 0.7285 | 0.7526 | 0.7289 | 0.7475 | 0.7273 | 0.7398 | 0.7186 |
| | (0.0040) | (0.0038) | (0.0041) | (0.0042) | (0.0042) | (0.0038) | (0.0034) | (0.0036) | (0.0037 | (0.0037) |
| Headcount | 0.0982 | 0.1153 | 0.0844 | 0.1149 | 0.0802 | 0.1130 | 0.0947 | 0.1197 | 0.1014 | 0.1323 |
| | (0.0066) | (0.0064) | (0.0066) | (0.0071) | (0.0060) | (0.0064) | (0.0059) | (0.0061) | (0.0061) | (0.0066) |
| $FGT_1$ | 0.0237 | 0.0285 | 0.0203 | 0.0269 | 0.0173 | 0.0264 | 0.0222 | 0.0301 | 0.0246 | 0.0336 |
| | (0.0023) | (0.0023) | (0.0021) | (0.0024) | (0.0020) | (0.0022) | (0.0020) | (0.0022) | (0.0020) | (0.0024) |
| $FGT_2$ | 0.0100 | 0.0122 | 0.0086 | 0.0112 | 0.0068 | 0.0105 | 0.0091 | 0.0129 | 0.0102 | 0.0145 |
| | (0.0014) | (0.0013) | (0.0012) | (0.0014) | (0.0012) | (0.0013) | (0.0011) | (0.0013) | (0.0011) | (0.0014) |

*Note*: The values are computed from the Beta mixture model.

**Table S2: First Order Stochastic Dominance Probabilities for Male Population**

| | *A* 2017 *B* 2014 | *A* 2014 *B* 2010 | *A* 2010 *B* 2006 | *A* 2006 *B* 2001 |
|---|---|---|---|---|
| Pr($A_{FSD}B$) | 0.0011 | 0.0005 | 0.0107 | 0.0392 |
| Pr($B_{FSD}A$) | 0.0170 | 0.0926 | 0.0126 | 0.0007 |
| Pr(no dominance) | 0.9819 | 0.9068 | 0.9767 | 0.9601 |
| | | 2017 2010 | 2014 2006 | 2010 2001 |
| Pr($A_{FSD}B$) | | 0.0000 | 0.0010 | 0.0094 |
| Pr($B_{FSD}A$) | | 0.0782 | 0.0962 | 0.0002 |
| Pr(no dominance) | | 0.9218 | 0.9028 | 0.9904 |
| | | | 2017 2006 | 2014 2001 |
| Pr($A_{FSD}B$) | | | 0.0001 | 0.0015 |
| Pr($B_{FSD}A$) | | | 0.1176 | 0.0090 |
| Pr(no dominance) | | | 0.8823 | 0.9895 |
| | | | | 2017 2001 |
| Pr($A_{FSD}B$) | | | | 0.0010 |
| Pr($B_{FSD}A$) | | | | 0.0557 |
| Pr(no dominance) | | | | 0.9433 |

**Table S3: First Order Stochastic Dominance Probabilities for Female Population**

| | *A* 2017 *B* 2014 | *A* 2014 *B* 2010 | *A* 2010 *B* 2006 | *A* 2006 *B* 2001 |
|---|---|---|---|---|
| Pr($A_{FSD}B$) | 0.0010 | 0.0027 | 0.0043 | 0.0212 |
| Pr($B_{FSD}A$) | 0.0457 | 0.0362 | 0.0183 | 0.0015 |
| Pr(no dominance) | 0.9533 | 0.9611 | 0.9774 | 0.9773 |
| | | 2017 2010 | 2014 2006 | 2010 2001 |
| Pr($A_{FSD}B$) | | 0.0001 | 0.0007 | 0.0066 |
| Pr($B_{FSD}A$) | | 0.0748 | 0.0522 | 0.0029 |
| Pr(no dominance) | | 0.9251 | 0.9471 | 0.9905 |
| | | | 2017 2006 | 2014 2001 |
| Pr($A_{FSD}B$) | | | 0.0001 | 0.0020 |
| Pr($B_{FSD}A$) | | | 0.1535 | 0.0113 |
| Pr(no dominance) | | | 0.8464 | 0.9867 |
| | | | | 2017 2001 |
| Pr($A_{FSD}B$) | | | | 0.0003 |
| Pr($B_{FSD}A$) | | | | 0.0767 |
| Pr(no dominance) | | | | 0.9230 |

**Table S4: Second Order Stochastic Dominance Probabilities for Male Population**

| | *A* 2017 – *B* 2014 | *A* 2014 – *B* 2010 | *A* 2010 – *B* 2006 | *A* 2006 – *B* 2001 |
|---|---|---|---|---|
| $\Pr(A_{SSD}B)$ | 0.0141 | 0.0097 | 0.3439 | 0.4428 |
| $\Pr(B_{SSD}A)$ | 0.3792 | 0.4740 | 0.0562 | 0.0216 |
| Pr(no dominance) | 0.6067 | 0.5163 | 0.5999 | 0.5356 |
| | | 2017 – 2010 | 2014 – 2006 | 2010 – 2001 |
| $\Pr(A_{SSD}B)$ | | 0.0002 | 0.0649 | 0.7065 |
| $\Pr(B_{SSD}A)$ | | 0.5905 | 0.2457 | 0.0026 |
| Pr(no dominance) | | 0.4093 | 0.6894 | 0.2909 |
| | | | 2017 – 2006 | 2014 – 2001 |
| $\Pr(A_{SSD}B)$ | | | 0.0035 | 0.3727 |
| $\Pr(B_{SSD}A)$ | | | 0.2822 | 0.0537 |
| Pr(no dominance) | | | 0.7143 | 0.5736 |
| | | | | 2017 – 2001 |
| $\Pr(A_{SSD}B)$ | | | | 0.0578 |
| $\Pr(B_{SSD}A)$ | | | | 0.1523 |
| Pr(no dominance) | | | | 0.7899 |

**Table S5: Second Order Stochastic Dominance Probabilities for Female Population**

| | *A* 2017 – *B* 2014 | *A* 2014 – *B* 2010 | *A* 2010 – *B* 2006 | *A* 2006 – *B* 2001 |
|---|---|---|---|---|
| $\Pr(A_{SSD}B)$ | 0.0120 | 0.0169 | 0.2090 | 0.3946 |
| $\Pr(B_{SSD}A)$ | 0.4603 | 0.4441 | 0.1027 | 0.0321 |
| Pr(no dominance) | 0.5277 | 0.5390 | 0.6883 | 0.5733 |
| | | 2017 – 2010 | 2014 – 2006 | 2010 – 2001 |
| $\Pr(A_{SSD}B)$ | | 0.0006 | 0.0428 | 0.5058 |
| $\Pr(B_{SSD}A)$ | | 0.8288 | 0.3829 | 0.0250 |
| Pr(no dominance) | | 0.1706 | 0.5743 | 0.4692 |
| | | | 2017 – 2006 | 2014 – 2001 |
| $\Pr(A_{SSD}B)$ | | | 0.0012 | 0.1328 |
| $\Pr(B_{SSD}A)$ | | | 0.7741 | 0.1300 |
| Pr(no dominance) | | | 0.2247 | 0.7372 |
| | | | | 2017 – 2001 |
| $\Pr(A_{SSD}B)$ | | | | 0.0133 |
| $\Pr(B_{SSD}A)$ | | | | 0.4596 |
| Pr(no dominance) | | | | 0.5271 |

**Table S6: First Order Stochastic Dominance Probabilities for Male Population with a Mental Health Score below 0.5**

| | *A* 2017 | *B* 2014 | *A* 2014 | *B* 2010 | *A* 2010 | *B* 2006 | *A* 2006 | *B* 2001 |
|---|---|---|---|---|---|---|---|---|
| Pr($A_{FSD}B$) | 0.0325 | | 0.0047 | | 0.4393 | | 0.4293 | |
| Pr($B_{FSD}A$) | 0.3262 | | 0.5259 | | 0.0398 | | 0.0162 | |
| Pr(no dominance) | 0.6413 | | 0.4694 | | 0.5209 | | 0.5545 | |
| | | | 2017 | 2010 | 2014 | 2006 | 2010 | 2001 |
| Pr($A_{FSD}B$) | | | 0.0004 | | 0.0554 | | 0.7539 | |
| Pr($B_{FSD}A$) | | | 0.5854 | | 0.2382 | | 0.0012 | |
| Pr(no dominance) | | | 0.4152 | | 0.7064 | | 0.2449 | |
| | | | | | 2017 | 2006 | 2014 | 2001 |
| Pr($A_{FSD}B$) | | | | | 0.0054 | | 0.4014 | |
| Pr($B_{FSD}A$) | | | | | 0.2685 | | 0.0460 | |
| Pr(no dominance) | | | | | 0.7261 | | 0.5526 | |
| | | | | | | | 2017 | 2001 |
| Pr($A_{FSD}B$) | | | | | | | 0.1624 | |
| Pr($B_{FSD}A$) | | | | | | | 0.1162 | |
| Pr(no dominance) | | | | | | | 0.7214 | |

**Table S7: First Order Stochastic Dominance Probabilities for Female Population with a Mental Health Score below 0.5**

| | *A* 2017 | *B* 2014 | *A* 2014 | *B* 2010 | *A* 2010 | *B* 2006 | *A* 2006 | *B* 2001 |
|---|---|---|---|---|---|---|---|---|
| Pr($A_{FSD}B$) | 0.0230 | | 0.0116 | | 0.1771 | | 0.2819 | |
| Pr($B_{FSD}A$) | 0.4215 | | 0.5409 | | 0.0853 | | 0.0501 | |
| Pr(no dominance) | 0.5555 | | 0.4475 | | 0.7376 | | 0.6680 | |
| | | | 2017 | 2010 | 2014 | 2006 | 2010 | 2001 |
| Pr($A_{FSD}B$) | | | 0.0007 | | 0.0228 | | 0.3738 | |
| Pr($B_{FSD}A$) | | | 0.8254 | | 0.4235 | | 0.0447 | |
| Pr(no dominance) | | | 0.1739 | | 0.5537 | | 0.5815 | |
| | | | | | 2017 | 2006 | 2014 | 2001 |
| Pr($A_{FSD}B$) | | | | | 0.0012 | | 0.0831 | |
| Pr($B_{FSD}A$) | | | | | 0.7518 | | 0.3220 | |
| Pr(no dominance) | | | | | 0.2470 | | 0.5949 | |
| | | | | | | | 2017 | 2001 |
| Pr($A_{FSD}B$) | | | | | | | 0.0074 | |
| Pr($B_{FSD}A$) | | | | | | | 0.4964 | |
| Pr(no dominance) | | | | | | | 0.4962 | |

**Table S8: Second Order Stochastic Dominance Probabilities for Male Population with a Mental Health Score below 0.5**

| | *A* 2017 | *B* 2014 | *A* 2014 | *B* 2010 | *A* 2010 | *B* 2006 | *A* 2006 | *B* 2001 |
|---|---|---|---|---|---|---|---|---|
| Pr($A_{SSD}B$) | 0.0925 | | 0.0212 | | 0.5915 | | 0.4909 | |
| Pr($B_{SSD}A$) | 0.3963 | | 0.5537 | | 0.0760 | | 0.0749 | |
| Pr(no dominance) | 0.5112 | | 0.4251 | | 0.3325 | | 0.4342 | |
| | | | 2017 | 2010 | 2014 | 2006 | 2010 | 2001 |
| Pr($A_{SSD}B$) | | | 0.0023 | | 0.1615 | | 0.7763 | |
| Pr($B_{SSD}A$) | | | 0.5934 | | 0.2852 | | 0.0051 | |
| Pr(no dominance) | | | 0.4043 | | 0.5533 | | 0.2186 | |
| | | | | | 2017 | 2006 | 2014 | 2001 |
| Pr($A_{SSD}B$) | | | | | 0.0378 | | 0.5305 | |
| Pr($B_{SSD}A$) | | | | | 0.2845 | | 0.0956 | |
| Pr(no dominance) | | | | | 0.6777 | | 0.3739 | |
| | | | | | | | 2017 | 2001 |
| Pr($A_{SSD}B$) | | | | | | | 0.3253 | |
| Pr($B_{SSD}A$) | | | | | | | 0.1617 | |
| Pr(no dominance) | | | | | | | 0.5130 | |

**Table S9: Second Order Stochastic Dominance Probabilities for Female Population with a Mental Health Score below 0.5**

| | *A* 2017 | *B* 2014 | *A* 2014 | *B* 2010 | *A* 2010 | *B* 2006 | *A* 2006 | *B* 2001 |
|---|---|---|---|---|---|---|---|---|
| Pr($A_{SSD}B$) | 0.0732 | | 0.0230 | | 0.3208 | | 0.4973 | |
| Pr($B_{SSD}A$) | 0.4759 | | 0.7234 | | 0.1656 | | 0.0921 | |
| Pr(no dominance) | 0.4509 | | 0.2536 | | 0.5136 | | 0.4106 | |
| | | | 2017 | 2010 | 2014 | 2006 | 2010 | 2001 |
| Pr($A_{SSD}B$) | | | 0.0019 | | 0.0559 | | 0.5821 | |
| Pr($B_{SSD}A$) | | | 0.8461 | | 0.6060 | | 0.0703 | |
| Pr(no dominance) | | | 0.1520 | | 0.3381 | | 0.3476 | |
| | | | | | 2017 | 2006 | 2014 | 2001 |
| Pr($A_{SSD}B$) | | | | | 0.0051 | | 0.1485 | |
| Pr($B_{SSD}A$) | | | | | 0.7967 | | 0.4321 | |
| Pr(no dominance) | | | | | 0.1982 | | 0.4194 | |
| | | | | | | | 2017 | 2001 |
| Pr($A_{SSD}B$) | | | | | | | 0.0335 | |
| Pr($B_{SSD}A$) | | | | | | | 0.5279 | |
| Pr(no dominance) | | | | | | | 0.4386 | |

**Table S10: Posterior Means (Standard Deviations) for Mean Health Scores for Indigenous and Non-indigenous Subgroups**

| | 2001 | | 2006 | | 2010 | | 2014 | | 2017 | |
|---|---|---|---|---|---|---|---|---|---|---|
| | Indig | Non-Ind | Indig | Non-Ind | Indig | Non-Ind | Indig | Non-Ind | Indig | Non-Ind |
| Mean | 0.7048 | 0.7388 | 0.6989 | 0.7475 | 0.7023 | 0.7466 | 0.6644 | 0.7388 | 0.6708 | 0.7295 |
| | (0.0208) | (0.0032) | (0.0232) | (0.0033) | (0.0193) | (0.0032) | (0.0199) | (0.0030) | (0.0168) | (0.0031) |
| Headcount | 0.1416 | 0.1016 | 0.1628 | 0.0923 | 0.1347 | 0.0945 | 0.2313 | 0.1054 | 0.1893 | 0.1190 |
| | (0.0352) | (0.0051) | (0.0386) | (0.0056) | (0.0320) | (0.0051) | (0.0353) | (0.0049) | (0.0307) | (0.0052) |
| $FGT_1$ | 0.0348 | 0.0253 | 0.0442 | 0.0222 | 0.0324 | 0.0212 | 0.0718 | 0.0267 | 0.0484 | 0.0300 |
| | (0.0121) | (0.0017) | (0.0149) | (0.0019) | (0.0105) | (0.0016) | (0.0156) | (0.0016) | (0.0118) | (0.0018) |
| $FGT_2$ | 0.0150 | 0.0109 | 0.0206 | 0.0094 | 0.0142 | 0.0083 | 0.0348 | 0.0114 | 0.0212 | 0.0129 |
| | (0.0073) | (0.0011) | (0.0094) | (0.0011) | (0.0064) | (0.0009) | (0.0098) | (0.0010) | (0.0071) | (0.0011) |

*Note*: The values are computed from the Beta mixture model.

**Table S11: First Order Stochastic Dominance Probabilities for Indigenous Population**

| | *A* 2017 – *B* 2014 | *A* 2014 – *B* 2010 | *A* 2010 – *B* 2006 | *A* 2006 – *B* 2001 |
|---|---|---|---|---|
| Pr($A_{FSD}B$) | 0.0708 | 0.0012 | 0.0549 | 0.0452 |
| Pr($B_{FSD}A$) | 0.0092 | 0.1984 | 0.0454 | 0.0705 |
| Pr(no dominance) | 0.9200 | 0.8004 | 0.8997 | 0.8843 |
| | | 2017 – 2010 | 2014 – 2006 | 2010 – 2001 |
| Pr($A_{FSD}B$) | | 0.0071 | 0.0026 | 0.0446 |
| Pr($B_{FSD}A$) | | 0.1941 | 0.2411 | 0.0717 |
| Pr(no dominance) | | 0.7988 | 0.7563 | 0.8837 |
| | | | 2017 – 2006 | 2014 – 2001 |
| Pr($A_{FSD}B$) | | | 0.0109 | 0.0003 |
| Pr($B_{FSD}A$) | | | 0.1630 | 0.2755 |
| Pr(no dominance) | | | 0.8261 | 0.7242 |
| | | | | 2017 – 2001 |
| Pr($A_{FSD}B$) | | | | 0.0063 |
| Pr($B_{FSD}A$) | | | | 0.2269 |
| Pr(no dominance) | | | | 0.7668 |

**Table S12: Second Order Stochastic Dominance Probabilities for Indigenous Population**

| | *A* 2017 – *B* 2014 | *A* 2014 – *B* 2010 | *A* 2010 – *B* 2006 | *A* 2006 – *B* 2001 |
|---|---|---|---|---|
| Pr($A_{SSD}B$) | 0.4381 | 0.0097 | 0.3377 | 0.1254 |
| Pr($B_{SSD}A$) | 0.0524 | 0.7298 | 0.1204 | 0.3788 |
| Pr(no dominance) | 0.5095 | 0.2605 | 0.5419 | 0.4958 |
| | | 2017 2010 | 2014 2006 | 2010 2001 |
| Pr($A_{SSD}B$) | | 0.0399 | 0.0469 | 0.2017 |
| Pr($B_{SSD}A$) | | 0.5255 | 0.5469 | 0.2491 |
| Pr(no dominance) | | 0.4346 | 0.4062 | 0.5492 |
| | | | 2017 2006 | 2014 2001 |
| Pr($A_{SSD}B$) | | | 0.1222 | 0.0053 |
| Pr($B_{SSD}A$) | | | 0.3169 | 0.7705 |
| Pr(no dominance) | | | 0.5609 | 0.2242 |
| | | | | 2017 2001 |
| Pr($A_{SSD}B$) | | | | 0.0400 |
| Pr($B_{SSD}A$) | | | | 0.5248 |
| Pr(no dominance) | | | | 0.4352 |

**Table S13: First Order Stochastic Dominance Probabilities for Indigenous Population with a Mental Health Score below 0.5**

| | *A* 2017 *B* 2014 | *A* 2014 *B* 2010 | *A* 2010 *B* 2006 | *A* 2006 *B* 2001 |
|---|---|---|---|---|
| Pr($A_{FSD}B$) | 0.5774 | 0.0064 | 0.4115 | 0.1314 |
| Pr($B_{FSD}A$) | 0.0360 | 0.7673 | 0.1144 | 0.4239 |
| Pr(no dominance) | 0.3866 | 0.2263 | 0.4741 | 0.4447 |
| | | 2017 2010 | 2014 2006 | 2010 2001 |
| Pr($A_{FSD}B$) | | 0.0456 | 0.0424 | 0.2523 |
| Pr($B_{FSD}A$) | | 0.5246 | 0.5731 | 0.2572 |
| Pr(no dominance) | | 0.4298 | 0.3845 | 0.4905 |

| | | | 2017 2006 | 2014 2001 |
|---|---|---|---|---|
| Pr($A_{FSD} B$) | | | 0.1906 | 0.0045 |
| Pr($B_{FSD} A$) | | | 0.2908 | 0.7864 |
| Pr(no dominance) | | | 0.5186 | 0.2091 |
| | | | | 2017 2001 |
| Pr($A_{FSD} B$) | | | | 0.0612 |
| Pr($B_{FSD} A$) | | | | 0.5007 |
| Pr(no dominance) | | | | 0.4381 |

**Table S14: Second Order Stochastic Dominance Probabilities for Indigenous Population with a Mental Health Score below 0.5**

| | *A* 2017 *B* 2014 | *A* 2014 *B* 2010 | *A* 2010 *B* 2006 | *A* 2006 *B* 2001 |
|---|---|---|---|---|
| Pr($A_{SSD} B$) | 0.6554 | 0.0157 | 0.4769 | 0.1901 |
| Pr($B_{SSD} A$) | 0.0632 | 0.7814 | 0.1876 | 0.5087 |
| Pr(no dominance) | 0.2814 | 0.2029 | 0.3355 | 0.3012 |
| | | 2017 2010 | 2014 2006 | 2010 2001 |
| Pr($A_{SSD} B$) | | 0.1033 | 0.0833 | 0.3230 |
| Pr($B_{SSD} A$) | | 0.5599 | 0.6120 | 0.3468 |
| Pr(no dominance) | | 0.3368 | 0.3047 | 0.3302 |
| | | | 2017 2006 | 2014 2001 |
| Pr($A_{SSD} B$) | | | 0.3048 | 0.0124 |
| Pr($B_{SSD} A$) | | | 0.3488 | 0.8084 |
| Pr(no dominance) | | | 0.3464 | 0.1792 |
| | | | | 2017 2001 |
| Pr($A_{SSD} B$) | | | | 0.1305 |

| | |
|---|---|
| Pr($B_{SSD}A$) | 0.5514 |
| Pr(no dominance) | 0.3181 |

**C. References**